\documentclass[10pt,journal,compsoc]{IEEEtran}

\ifCLASSOPTIONcompsoc
  \usepackage[nocompress]{cite}
\else
  \usepackage{cite}
\fi

\usepackage{cite} 
\usepackage{amsmath,amssymb,amsfonts}
\usepackage{algorithmic}
\usepackage{graphicx}
\usepackage{textcomp,lineno,hyperref,chngcntr,caption,wasysym,textcomp,mathrsfs,indentfirst,subfig,verbatim,url} 
\usepackage{verbatim}
\usepackage{algorithm}
\usepackage{algorithmic}
\usepackage{diagbox}
\usepackage{balance}

\begin{document}

\title{Unexpected Information Leakage of Differential Privacy Due to Linear Property of Queries}

\author{Wen Huang,Shijie Zhou,Yongjian Liao
\IEEEcompsocitemizethanks{\IEEEcompsocthanksitem Wen Huang, Shijie Zhou and Yongjian Liao is with the School of Information and Software Engineering,University of Electronic Science and Technology of China,Chengdu, Sichuan, China.\protect\\
E-mail: 562421007@qq.com}
\thanks{Manuscript received October 17, 2020; revised October 17, 2020.}}

%
%

\markboth{Journal of \LaTeX\ Class Files,~Vol.~14, No.~8, August~2015}%
{Shell \MakeLowercase{\textit{et al.}}: Bare Advanced Demo of IEEEtran.cls for IEEE Computer Society Journals}
%

\IEEEtitleabstractindextext{%

\begin{abstract}
The differential privacy is a widely accepted conception of privacy preservation and the Laplace mechanism is a famous instance of differential privacy mechanisms to deal with numerical data. In this paper, we find that the differential privacy does not take liner property of queries into account, resulting in unexpected information leakage. In specific, the linear property makes it possible to divide one query into two queries such as $q(D)=q(D_1)+q(D_2)$ if $D=D_1\cup D_2$ and $D_1\cap D_2=\emptyset$. If attackers try to obtain an answer of $q(D)$, they not only can issue the query $q(D)$, but also can issue the $q(D_1)$ and calculate the $q(D_2)$ by themselves as long as they know $D_2$. By different divisions of one query, attackers can obtain multiple different answers for the query from  differential privacy mechanisms. However, from attackers' perspective and from differential privacy mechanisms' perspective, the totally consumed privacy budget is different if divisions are delicately designed. The difference leads to unexpected information leakage because the privacy budget is the key parameter to control the amount of legally released information from differential privacy mechanisms. In order to demonstrate the unexpected information leakage, we present a membership inference attacks against the Laplace mechanism. Specifically, we propose a method to obtain multiple independent identical distribution samples of linear query's answer under constrains of differential privacy. The proposed method is based on linear property and some background knowledge of attackers. When the background knowledge is enough, the proposed method can obtain enough number of samples from differential privacy mechanisms such that the totally consumed privacy budget can be unreasonably large.  Based on obtained samples, a hypothesis test method is used to determine whether a target record is in a target data set.
\end{abstract}

}

\maketitle

\IEEEdisplaynontitleabstractindextext

\IEEEpeerreviewmaketitle

\ifCLASSOPTIONcompsoc
\IEEEraisesectionheading{\section{Introduction}\label{sec:introduction}}
\else
\section{Introduction}
\label{sec:introduction}
\fi

\par The differential privacy is state of the art conception for privacy preservation because it formalizes a strong privacy guarantee by a solid mathematical foundation. That is, even if there is only one different record in two data sets, it is hard to tell one data set from another data set. In differential privacy, the privacy guarantee is quantified by probability with which attackers can tell one data set from another data set. The probability is controlled by  a parameter called privacy budget.  

\par One of most important reasons for popularity of differential privacy is that it makes no assumptions about the background knowledge available to attackers  \cite{29}. That is, even if attackers know all data records except one, attackers cannot get much information of the record which attackers don't know. The amount of information which attackers can extract is strictly controlled by the privacy budget. However, intuitively, if attackers have more background knowledge, attackers have stronger ability to extract information. For example, if attackers know a target record related to a female, attackers can filter all records related to male, resulting in high probability to get target information.  

\par In this paper, we explore the transformation from background knowledge to  consuming extra privacy budget of differential privacy mechanisms. That is, how attackers can consume unexpected privacy budget by its background knowledge. We find out that the transformation is possible because differential privacy mechanisms do not take one property of query into account, namely linear property. 

\par The linear property makes it possible to divide one query into two queries. For example, the counting query is one class of linear queries. For query $count(\{b_1,b_2,b_3\})$, the next equality holds.
\begin{eqnarray*}
	& & count(\{b_1,b_2,b_3\} \\
	&=& count(\{b_1,b_2 \})+count(\{ b_3\})\\
	&=&count(\{b_1,b_3 \})+count(\{ b_2\})\\	
\end{eqnarray*}

\par Suppose that attackers aim at the answer of $count(\{b_1,b_2,b_3\})$ and the consumed privacy budget is $\epsilon$ when a differential privacy mechanism responds one query every time.  

\par Attackers can issue two different queries, namely $count(\{b_1,b_2 \})$ and $count(\{b_1,b_3 \})$, to a differential privacy mechanism and obtain two responses. From the differential privacy mechanism's perspective, it answers two different queries and each query consumes privacy budget $\epsilon$. 

\par However, if $\{b_2,b_3\}$ is background knowledge of attackers, attackers can get two answers for the same query. In specific, by adding up the first response which is an answer of $count(\{b_1,b_2 \})$ and $count(\{b_3\})$ which is computed by attackers themselves, attackers can obtain an answer of $count(\{b_1,b_2,b_3\}$. By adding up the second response and $count(\{b_2\})$, attackers can obtain another answer of $count(\{b_1,b_2,b_3\}$. So, from attackers'  perspective, the differential privacy mechanism answers the same query two times and consumed privacy budget for each time is $\epsilon$. That is, the totally consumed privacy budget is $2*\epsilon$ for the same query.   

\par The possibility to obtain multiple answers for one query leads to risk of leaking membership information of records. In specific, attackers obtain multiple answers of query $count(\{b_1,b_2,b_3\})$ and they know that the record $b_2$ and $b_3$ are in the target data set. If the record $b_1$ is in the target data set, the mean of obtained samples is 3. Otherwise, the mean of samples is 2. By estimating the mean of samples, attackers can determine whether the record $b_1$ is in the target data set. 

\par In a word, if the background knowledge of attackers is enough, attackers can consume unreasonably large privacy budget in total while differential privacy mechanisms don't beware, resulting in risk of information leakage. The risk may be a big challenge for applications of differential privacy mechanisms because influenced linear query such as the counting query and the sum query are elemental queries which differential privacy mechanisms can handle well in previous researches. 

\par In order to demonstrate the unexpected information leakage of differential privacy mechanisms, we take advantage of the above vulnerability to construct a membership inference attacks method against the Laplace mechanism. Our main contributions are as follows  

\par (1) We point out that the liner property of queries is a source of information leakage for differential privacy. To the best of our knowledge, this paper is the first one to discuss information leakage caused by linear property of queries. 

\par (2) A method is proposed to obtain multiple i.i.d.(independent identical distribution) samples for linear query's answer under constrains of differential privacy. In general cases, multiple i.i.d. samples for query's answer cannot be obtained directly in differential privacy. However, the linear query has two special properties. The first property is linear property and the second property is that the sensitivity can be calculated easily. Based on these two properties, we propose a method to obtain multiple i.i.d. samples for linear query's answer.


\par (3) A membership inference attacks method against the Laplace mechanism is proposed. The goal of membership inference attacks is to determine whether a target record is in a target data set. However, the differential privacy claims that even if attackers know all records except one, attackers cannot extract information of the record they don't know. Therefore, the membership inference attacks is an appropriate way to demonstrate the information leakage of differential privacy. In the proposed attacks method, the decision is made by hypothesis test method which is based on multiple i.i.d. samples obtained.


\subsection{Related Work}

\par The differential privacy is state of the art conception for privacy preservation \cite{26}. Since the conception of differential privacy is proposed, there have been some analyses about its information leakage. There are mainly two lines of papers. The first line is related to assumption of data generation. In differential privacy, records in data sets are assumed to be independent. But the assumption is not always true in real application scenarios \cite{31}. For example, Bob or one of his 9 immediate family members may have contracted a highly infectious disease. The whole family is infected together or none of them is infected with high probability \cite{32}. The correlation among family members makes it easier for attackers to infer whether Bob is infected by querying "how many people are infected by the disease in Bob's family". There are some papers aiming at fixing the correlation problem.  Lv et al. research the problem in scenarios of big data \cite{33}.  Wu et al. research the problem from the perspective of game theory \cite{19}. Zhang et al. research the problem in machine learning \cite{34}. The key of fixing correlation problem is to find an appropriate quantity to quantify correlation among data records. Then added noise of differential privacy mechanisms is calibrated by the quantity.  Chen et al. quantify degree of correlation by Gaussian correlation model \cite{35}. Zhu et al. propose conception of correlated sensitivity to quantify degree of correlation \cite{24}. A strength of the line is that the unexpected information leakage is not related to certain differential privacy mechanisms. A weakness is that the unexpected information leakage is only related to correlated records not for all records. 
\par The second line is related to the extensional conception of sensitivity of differential privacy. There is a difficult trade-off in differential privacy mechanisms, namely trade-off between magnitude of noise and data utility \cite{27} \cite{28}. In order to reduce the magnitude of noise, various alternative conceptions for sensitivity are proposed such as elastic sensitivity \cite{22} and record sensitivity \cite{26}. Some of these proposed conceptions are weak and lead to unexpected information leakage. The local sensitivity is an example \cite{23}. The goal of local sensitivity is to release data with database-based additive noise. That is, the noise magnitude is determined not only by the queries which are to be released, but also by the data set itself. However, the noise calibrated by local sensitivity is too small, resulting in information leakage. For example, let $f_{med}(x) = median(x_1,x_2,\dots,x_n)$, where $x_i$ are sorted real number from bound interval such as $[0,\Lambda]$. If the $f_{med}(x)$ is released with noise magnitude proportional to local sensitivity, then the probability of receiving a non-zero answer is zero when $x_1=\dots = x_{m+1}=0$,$x_{m+2}=\dots=x_n=\Lambda$. Whereas the probability of receiving a non-zero answer is non-negligible when $x_1=\dots = x_{m}=0$,$x_{m+1}=\dots=x_n=\Lambda$. Where $n$ is odd and $m=\frac{n+1}{2}$. The analyses for extensional conceptions of sensitivity are difficult because how big the noise magnitude needs to be is hard to quantify.


\par The goal of membership inference attacks is to determine that a given data record is in a target data set or not. The attack is firstly demonstrated by Homer  \cite{36}.  Their paper has so much big influence on privacy community that the USA National Institute of Health (NIH) switches policies in case of membership information leakage in process of releasing statistical data. The membership inference attacks has been valued since then. Shokri et al. quantitatively investigate membership inference attacks against machine learning models \cite{43}. Rahman et al. analyze membership inference attacks against deep learning model which is protected by differential privacy mechanisms \cite{41}.  Yeom et al. investigate the relationship between overfitting and membership inference attacks, finding that "overfitting is sufficient to allow an attacker to perform membership inference attack" \cite{42}. Backes et al. explore the membership inference attacks against miRNA expression data set \cite{44}. Almadhoun et al. discuss membership inference attacks against differential privacy mechanisms \cite{37}. They take advantage of dependence of data records. Almadhoun et al also analyze an application of membership inference against genomic data sets \cite{38}.    
\par In brief, the first line is about the special records which are correlated with each other and the second line is about trade-off between noise magnitude and data utility. Different from the mentioned two lines, we will discuss unexpected information leakage from a novel angle. In this paper, from the view of query functions, we show that differential privacy mechanisms do not take liner property of queries into account, resulting in unexpected information leakage.

\subsection{Organization}
\par In the next section, background knowledge will be introduced briefly, including the conception of differential privacy, the linear property and the hypothesis test. In the section 3, how to obtain multiple answers for linear queries is presented.  In the section 4, the proposed membership inference attacks method will be presented and an instance of attack will be given. In section 5, experimental results  will be showed. At last, the conclusion will be claimed.

\section{background}  
\par The introduction of background knowledge is divided into three parts including basic conceptions of differential privacy, the linear property and the hypothesis test method.

\subsection{Differential Privacy} 
\par The differential privacy is the most popular conception of privacy preservation. It formalizes a strong privacy guarantee: even if attackers know all data records except one, attackers cannot get much information of the record which attackers don't know \cite{30} \cite{39}. To that end, differential privacy mechanisms guarantee that its outputs are insensitive to any particular data record. That is, presence or absence of any record has limited influence on outputs of differential privacy mechanisms. The presence or absence of one record in a data set is captured by a conception of neighboring databases. Specifically, databases $D$ and $D'$ are neighboring databases denoted by $D \sim D'$ if there is only one different record between $D$ and $D'$ denoted by $d(D,D')=1$. 

\par \textbf{Definition 1}(differential privacy) Any random mechanism $M:D^n\to R^d$ preserves $\epsilon$-DP(differential privacy) if for any neighboring databases $D$ and $D'$ such that $d(D,D')=1$ and for all sets of possible output $S$:
\begin{eqnarray*}
	P\{M(D)\in S\} \le e^\epsilon P\{M(D')\in S\}
\end{eqnarray*}
\par Here, the $\epsilon$ is the privacy budget. The privacy guarantee is quantified by the privacy budget. The smaller the privacy budget is, the stronger the privacy guarantee is.

\par The Laplace mechanism is a famous and foundational mechanism in the field of differential privacy. It is to deal with numerical data. The Laplace mechanism is based on a conception of global sensitivity. The global sensitivity quantifies the max difference of query's answer, which is caused by presence or absence of one record.

\par \textbf{Definition 2} (global sensitivity) For a given database $D$ and query $q$, the global sensitivity denoted by $\Delta D$ is	
\begin{eqnarray*}
	\Delta D = \max \limits_{D \sim D'} |q(D)-q(D')|
\end{eqnarray*} 

\par The Laplace mechanism covers the difference by noise drawn from Laplace distribution with location parameter $\mu=0$ and scale parameter $b=\frac{\Delta D}{\epsilon}$.

\par \textbf{Definition 3} (Laplace mechanism) For given database $D$, query $q$ and privacy budget $\epsilon$, the output of Laplace mechanism is 
\begin{eqnarray*}
	M(q,D,\epsilon) = q(D) + Lap(0,\frac{\Delta D}{\epsilon}) 
\end{eqnarray*} 
\par Here, the $Lap(0,\frac{\Delta D}{\epsilon})$ represents noise drawn from a Laplace distribution with location parameter 0 and scale parameter $\frac{\Delta D}{\epsilon}$. 

\par Differential privacy mechanisms have many good properties which make it possible to build complex differential privacy mechanisms by basic block algorithms. Two of them are Sequential Composition Theorem and Parallel Composition Theorem.       
\par \textbf{Sequential Composition Theorem} Let $A_1,A_2, \cdots, A_k$ be $k$ algorithms that satisfy $\epsilon_1$-DP,$\epsilon_2$-DP, $\cdots$,$\epsilon_k$-DP respectively. Publishing $t = <t_1,t_2,\cdots, t_k>$ satisfies $(\sum_{i=1}^{k}\epsilon_i)$-DP. Where, $t_1=A_1(D),t_2=A_2(t_1,D), t_3=A_3(<t_1,t_2,D>),\cdots, t_k=A_k(<t_1,t_2,t_3,\cdots,t_{k-1}>,D)$.   

\par \textbf{Parallel Composition Theorem} Let $A_1,A_2, \cdots, A_k$ be $k$ algorithms that satisfy $\epsilon_1$-DP,$\epsilon_2$-DP, $\dots$,$\epsilon_k$-DP respectively. Publishing $t = <t_1,t_2,\cdots, t_k>$ satisfies $(max_{i\in [1,2\cdots k]}\epsilon_i)$-DP. Where, $t_1=A_1(D_1),t_2=A_2(t_1,D_2), t_3=A_3(<t_1,t_2,D_3>),\cdots, t_k=A_k(<t_1,t_2,t_3,\cdots,t_{k-1}>,D_k)$ and $D_i \cap D_j = \emptyset $ for all $i\ne j$,    

\subsection{Linear Property}
\par The linear query is one kind of fundamental queries. It is used widely. For example, sum and counting query are popular linear queries. The linear query has a very good property, namely linear property.
\par \textbf{Definition 4} (linear property) For linear query $q$ and data set $D$ such that $D = D_1 \cup D_2$ and $ \varnothing = D_1 \cap D_2 $, we have  
\begin{eqnarray*}
	q(D) = q(D_1) +q(D_2) 
\end{eqnarray*}
\par The counting query is one kind of linear queries so it satisfies the linear property.  

\subsection{Hypothesis Test}
\par The hypothesis test is a statistic tool to determine whether fluctuation of experimental results is only due to randomness. Its main goal is to guarantee the confidence level of experimental conclusions. To that end, there are three steps in hypothesis test logically. Firstly, an assumption about experimental conclusions is made. The assumption needs to be contrary to conclusions which are achieved by experimental results. The assumption is called null hypothesis denoted by $H_0$. The opposite hypothesis of null hypothesis is called alternative hypothesis denoted by $H_1$. Secondly, a statistics is chosen. The statistics needs to be related to the assumption and its probability distribution function is known. Thirdly, the probability of experimental results is calculated through the known distribution. If the probability is smaller than a predetermined threshold, experimental conclusions are reliable. The procedure can guarantee that when the assumption which is contrary to experimental conclusions is right, the experimental results occur with extremely small probability. That is, the experimental results are not only due to randomness.    

\par The concrete steps of hypothesis test method are clear. Firstly, a predetermined probability threshold needs to be chosen and it is called significance level. Secondly, the experimental results' probability is calculated and the probability is called $p$ value.  At last, a decision is made. In specific, if the $p$ value is greater than the predetermined significance level, alternative hypothesis is accepted. Otherwise, the null hypothesis is accepted. 

\par The one-sample t-test is an useful hypothesis test method. It is used to determine whether samples come from a distribution with a certain mean.

\section{multiple i.i.d. samples for linear query's answer } 

\par This section presents how to obtain multiple i.i.d. samples for linear query's answer under constrains of differential privacy. Firstly, the idea of proposed method will be introduced. Then, the proposed method will be presented.    

\par Usually, multiple i.i.d. samples for a query's answer cannot be obtained directly under constrains of differential privacy. Differential privacy mechanisms are random mechanisms so it can output different answers for the same query when the same query comes every time. However, in order to reduce consumption of the privacy budget, differential privacy mechanisms output the same answer for queries which have been answered before. To that end, there are some engineering methods available. For example, a table is maintained. The table records answered queries as well as answers of these queries. When a query comes, differential privacy mechanisms look up whether the query is in the table. If the query was answered before, the recorded answer will be returned so that differential privacy mechanisms do not need to compute the answer again. 


\par The linear property makes it possible to divide a linear query into two parts. For example, the counting query is a linear query. There are four numbers in a set  $B=\{b_1,b_2,b_3,b_4\}$. The equality $count(B)=count(\{b_1,b_2\})+count(\{b_3,b_4\})$ holds. In addition, there are multiple divisions for one query. For instance, the equality $count(B)=count(\{b_1\})+count(\{b_2,b_3,b_4\})$ also holds.    

\par The linear property makes it possible to obtain multiple different answers for one linear query under constrains of differential privacy. In specific, one linear query can be divided into two parts such as $count(B)=count(\{b_1,b_2\})+count(\{b_3,b_4\})$. If attackers want to obtain the answer of $count(B)$ under the condition that all queries are only calculated by a differential privacy mechanism, attackers can issue the query $count(B)$ to the differential privacy mechanism. In addition, in order to obtain another answer of $count(B)$, attackers can also issue the query $count(\{b_1,b_2\})$ to the differential privacy mechanism and calculate the query $count(\{b_3,b_4\})$ by himself when attackers know the set $\{b_3,b_4\}$. In brief, by multiple divisions of a query, it is possible to obtain multiple answers for a linear query under constrains of differential privacy. 

\par There is a necessary assumption for  attackers to obtain multiple answers for a linear query. In specific,  attackers need to know a set of records in the target data set which is protected by the differential privacy mechanism. The set of records can be regarded as background knowledge of attackers.   

\par The assumption about the background knowledge in the proposed method is weaker than the assumption in differential privacy. The differential privacy claimed that even if attackers know all data records except one, the record which attackers don't know can be protected \cite{30} \cite{39}. That is, the background knowledge of attackers could be all records except one in the differential privacy. In proposed method, the background knowledge is only a set of records instead of all records.

\subsection{Proposed Method}

\begin{figure*}[htbp]
	\centering
	\includegraphics[width=18.0cm]{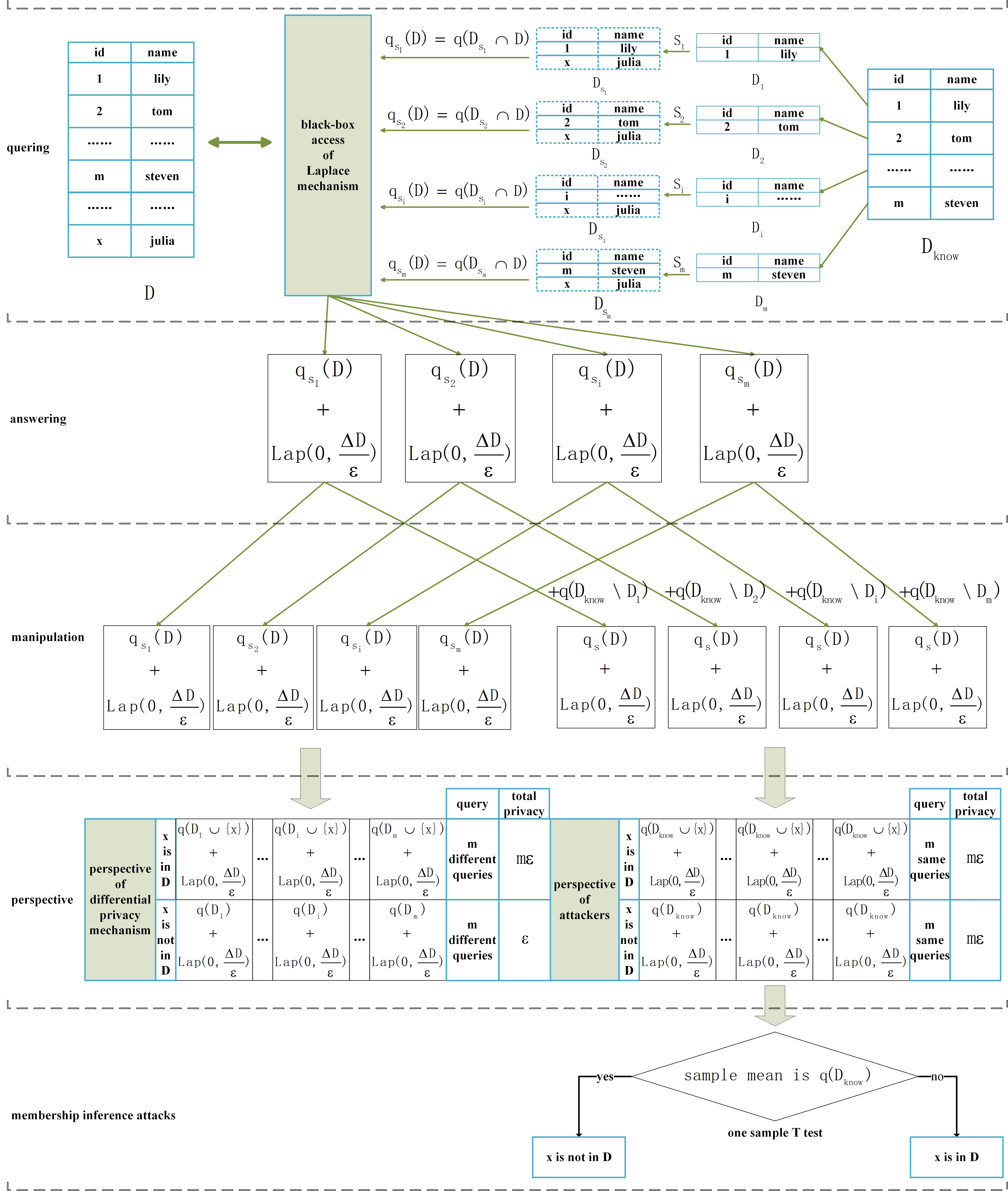}
	\caption*{\\ Fig 1: method to obtain multiple answers for the target query and membersip inference attacks mehtod} 
\end{figure*}

\par Attackers can issue query $q_s(D)$ to the differential privacy mechanism $M$. The $q_s(D)$ means computing the query $q$ over the target data set $D$ under the condition $s$ which specifies the range of data records. In other words, the $q_s(D)$ is equal to $q(D\cap D_s)$ where the $D_s$ is the set of records which satisfy the condition $s$. For example, "count the number of students whose age is greater than 10 in a classroom". The query $q$ is the counting query, the data set $D$ is the set of all students in the classroom and the condition $s$ is that " the age is greater than 10".

\par Suppose that the target record is denoted by $x$ and the background knowledge of attackers is a set of data records denoted by $D_{know}$. The goal is to obtain multiple answers for a linear query $q(\{x\}\cup D_{know})$. The $s$ represents the condition such that $D_s= \{x\}\cup D_{know}$. Therefore, the target query can be expressed as $q_s$ and its answer can be expressed as $M(q_s,D,\epsilon)$.

\par Attackers choose a subset $ D_i \subset D_{know}$, construct a query condition $s_i$ such that $D_{s_i} = D_i\cup \{x\}$, issue query $q_{s_i}(D)$ to the differential privacy mechanism and obtain $a_i = M(q_{s_i},D,\epsilon)= q_{s_i}(D)+Lap(0,\epsilon/\Delta D)$ as return. Let $\hat{a}_i = q(D_{know}\setminus D_i) + a_i$.
\par The data set $D_i$ is a subset of $D_{know}$ and the $D_{know}$ is background knowledge of  attackers so   attackers can compute $q(D_{know}\setminus D_i)$ locally by themselves.  The $\hat{a}_i$ is an answer of the target query $q(\{x\}\cup D_{know})$, which will be proved in next subsection. 

\par Through another subset $D_j \subset D_{know}$, attackers can get another answer of the target query. Therefore, attackers can get multiple answers of the target query. If $D_i \cap D_j = \emptyset$ for all $i\ne j$, the totally consumed privacy budget is different from the differential privacy mechanism's perspective and attackers' perspective. The claim is proved in next subsection. 
\par The whole idea of this paper is showed in Fig 1, including the method to obtain multiple answers for the target query and the membership inference attacks method which is presented in detail at next section.



\begin{algorithm}
	\caption*{method to obtain multiple answers of $q(\{x\}\cup D_{know})$}
	\label{alg:Framwork} 
	\begin{algorithmic}[1]
		\REQUIRE ~~\\ 
		the background knowledge  $D_{know}$;\\		
		the differential privacy mechanism $M$;\\
		a linear query $q$;\\
		the total privacy budget $\epsilon_t$;\\
		\ENSURE ~~\\
		the $m$ answers $\hat{a}_1,\hat{a}_2,\dots,\hat{a}_m$ ;\\	
		
		\STATE let $\epsilon = \frac{\epsilon_t}{m}$		
		\FOR{$i=1$ to $m$}
		\STATE choose a new subset $D_i \subset D_{know}$ \\ such that $D_i\cap D_j = \emptyset$ for all $j\ne i$ ;
		\STATE construct a query condition $s_i$ such that $D_{s_i} = D_i\cup \{x\}$;
		\STATE $a_i = M(q_{s_i},D,\epsilon)$;
		\STATE $\hat{a}_i = a_i + q(D_{know}\setminus D_i)$  	 
		\ENDFOR
		
		\RETURN $\hat{a}_1,\hat{a}_2,\dots,\hat{a}_m$		  
	\end{algorithmic}
\end{algorithm}



\subsection{Correctness and Privacy}
\par In this subsection, the proposed method is analyzed in terms of correctness and privacy respectively. In the theorem 1, we prove that linear queries' multiple answers can be obtained. In theorem 2, we analyze the amount of background knowledge of attackers. Then, totally consumed privacy budget is analyzed from the two perspectives. 


\par \textbf{Theorem 1}: $\hat{a}_1,\hat{a}_2,\dots,\hat{a}_m$ are i.i.d. samples of the target linear queries' answer calculated by differential privacy mechanisms over the target data set $D$.
\par \textbf{Proof}: 
\par For $\forall i$, we have    
\begin{eqnarray*}
	\hat{a}_i &=& q(D_{know}\setminus D_i) + a_i \\		
	\hat{a}_i &=& q(D_{know}\setminus D_i) + M(q_{s_i},D,\epsilon) \\
	\hat{a}_i &=& q(D_{know}\setminus D_i) + q_{s_i}(D) + Lap(0,\frac{\Delta D}{\epsilon})\\
\end{eqnarray*}

\par We know that $q_{s_i}(D) = q((D_i\cup \{x\})\cap D)$. And we also know 
\begin{eqnarray*}
	(D_i\cup \{x\})\cap D \ \subset \ D_i\cup \{x\}
\end{eqnarray*}
\par And
\begin{eqnarray*}
	(D_i\cup \{x\}) \ \cap \ (D_{know}\setminus D_i) = \emptyset
\end{eqnarray*}

\par In addition, $q$ is linear query. We have  

\begin{eqnarray*}	
	& & q(D_{know}\setminus D_i) + q_{s_i}(D)\\ 
	&=& q(D_{know}\setminus D_i) + q((D_i\cup \{x\})\cap D)   \\ 
	&=& q(D_u ) \\	
\end{eqnarray*}
\par Here $D_u$ is union set of $D_{know}\setminus D_i$ and $(D_i\cup \{x\})\cap D$.
\begin{eqnarray*}	
	D_u &=& (D_{know}\setminus D_i) \cup ((D_i\cup \{x\})\cap D) \\
	D_u &=& ((D_{know}\setminus D_i)\cap D) \cup ((D_i\cup \{x\})\cap D) \\	
	D_u &=& ((D_{know}\setminus D_i) \cup (D_i\cup \{x\}))\cap D) \\
	D_u &=& (D_{know} \cup  \{x\})\cap D) \\
\end{eqnarray*}
\par So, we have 
\begin{eqnarray*}
	q(D_u ) &=& q((D_{know} \cup  \{x\})\cap D )  \\
	q(D_u ) &=& q_s(D)  \\	
\end{eqnarray*}
\par So, we have
\begin{eqnarray*}
	\hat{a}_i &=& q(D_{know}\setminus D_i) + q_{s_i}(D) + Lap(0,\frac{\Delta D}{\epsilon})\\
	\hat{a}_i &=& q_s(D) + Lap(0,\frac{\Delta D}{\epsilon}) \ \ \ \ \ \ \ \ \ \ \ \ \ \ \ \ \ \ \ \ \ \ \ \ \ \ \ \ (1)\\
	\hat{a}_i &=& M(q_s,D,\epsilon)	
\end{eqnarray*}
\par According to the equality (1), for all $\hat{a}_1,\hat{a}_2,\dots,\hat{a}_m$, the noise $Lap(0,\frac{\Delta D}{\epsilon})$ are i.i.d. samples of a  Laplace distribution.  So, the $\hat{a}_1,\hat{a}_2,\dots,\hat{a}_m$ are i.i.d. samples of the target query's answer. 
$\hfill\Box$ 


\par The background knowledge is a foundation of proposed method. The background knowledge is a set of known records in the proposed method, although it is captured by prior probability distribution in some papers such as \cite{25}. It is easier to obtain some records of a target data set than to obtain a prior probability distribution of the target data set. For example, with respect to score of students, it is easier to collect some students' score than to obtain a prior probability distribution of  students' score.  

\par The number of records in $D_{know}$ is a way to quantify the amount of background knowledge of attackers.   

\par \textbf{Theorem 2}: The number of records in $D_{know}$ needs to be greater than $m$.
\par \textbf{Proof}: 
\par As known before, a subset of $D_{know}$ can  be used to generate an i.i.d. sample for the target query's answer. The number of i.i.d. samples is $m$. Therefore, the number of records in $D_{know}$ should guarantee that the number of possibly used subsets of $D_{know}$ is equal to or greater than $m$. 
\par In the proposed method, all possibly used subsets are disjoint with each other. Therefore, the number of possibly used subset is equal to the number of records in $D_{know}$ at most. In specific, each possibly used subset has only one record of $D_{know}$.
\par In a word, the number of records in $D_{know}$ needs to be greater than $m$. 
\par $\hfill\Box$  


\par The privacy budget is the key parameter for privacy guarantee in differential privacy mechanisms. The smaller the privacy budget is, the stronger the privacy guarantee is. One good property of differential privacy is that the totally consumed privacy budget can be calculated in a cumulative way.

\par \textbf{Theorem 3}: From attackers' perspective, the totally consumed privacy budget is $\epsilon_t$. 
\par \textbf{Proof}
\par For $\forall i$, the consumed privacy budget of $M(q_{s_i},D,\epsilon)$ is $\epsilon = \frac{\epsilon_t}{m}$. By the theorem 1, attackers obtain $m$ answers for the target query from the differential privacy mechanism. According to the Sequential Composition Theorem, the totally consumed privacy budget is     
\begin{eqnarray*}
	\sum_{i=1}^{m}\epsilon = m\epsilon = \epsilon_t
\end{eqnarray*}
$\hfill\Box$ 

\par From the attackers' perspective, the totally consumed privacy budget can be unreasonably large if there are enough records in the set of background knowledge. In specific, if attackers can consume $\epsilon$ privacy budget to obtain one answer of the target query, then attackers can consume $m*\epsilon$ privacy budget to obtain $m$ different answers of the target query using the proposed method. So, if there are enough records in the set of background knowledge, attackers can consume unreasonably large privacy budget for the target query. However, from the differential privacy mechanism's perspective, the totally consumed privacy budget is different.   

\par \textbf{Theorem 4}: From the differential privacy mechanism's perspective, when the target record $x$ is not in the target data set the totally consumed privacy budget is $\epsilon$. When the target record $x$ is in the target data set, the differential privacy mechanism responds $m$ different query with $\epsilon$ privacy budget for each response, in resulting  $\epsilon_t$ in total. 
\par \textbf{Proof}
\par Firstly, when the target record $x$ is not in the target data set $D$, for $\forall\ i$ we have
\begin{eqnarray*}
	q_{s_i} = q((D_i \cup \{x\})\cap D)	= q(D_i\cap D)=q(D_i)
\end{eqnarray*}
\par The consumed privacy budget of $q_{s_i}$ is denoted by $\epsilon_i$. For $\forall\ 0<i \ne j < m$, we know  $D_i \cap D_j = \emptyset$. According to the Parallel Composition Theorem, the totally consumed privacy budget is
\begin{eqnarray*}
	\max\limits_{0<i<m} \epsilon_i = \epsilon
\end{eqnarray*}

\par Secondly, when the target record $x$ is in the target data set $D$, for $\forall\ i$ we have 
\begin{eqnarray*}
	q_{s_i} = q((D_i \cup \{x\})\cap D)	= q(D_i \cup \{x\})
\end{eqnarray*}
\par  For $\forall\ 0<i \ne j < m$, we have
\begin{eqnarray*} 
	(D_i\cup\{x\}) \cap (D_j\cup\{x\}) = \{x\} \ne \emptyset
\end{eqnarray*}

\par According to the Sequential Composition Theorem, the totally consumed privacy budget is
\begin{eqnarray*}
	\sum_{i=1}^{m}\epsilon_i = \sum_{i=1}^{m}\epsilon = m\epsilon = \epsilon_t
\end{eqnarray*}
\par In a word, from the differential privacy mechanism's perspective, it responds $m$ different query with $\epsilon$ privacy budget for each response, in resulting  $\epsilon_t$ in total. That is, from the differential privacy mechanism's perspective, the target query only consume privacy budget $\epsilon$.
$\hfill\Box$    

\par In a word, the totally consumed privacy budget is delicately analyzed from attackers' perspective and differential privacy mechanism's perspective in the theorem 3 and 4 respectively. The attackers can consume unreasonably large privacy budget while the differential privacy mechanism may not beware. Whether the differential privacy mechanism bewares the unreasonably large consumed privacy budget depends on weather the target record is in the target data set. That is, the presence or absence of the target record has key influence on behavior of the differential privacy mechanism. Thus, it is possible to infer whether the target record is in the target data set by observing outputs of differential privacy mechanism.   



\section{membership inference attacks method}

\par In order to demonstrate the unexpected information leakage of differential privacy mechanisms due to linear property, a membership inference attacks method against the Laplace mechanism is constructed based on linear queries. 

\par In this section, the main content is divided into five subsections. In the first subsection, the membership inference attacks model is introduced. In the second subsection, an analysis is given to discuss the security foundation of the Laplace mechanism. In the third subsection, reasons why counting query is an appropriate liner query to perform membership inference attacks is discussed. In the fourth subsection, the membership inference attacks method is presented. In the last subsection, the success rate of proposed attacks method is analyzed. 

\subsection{Membership Inference Attack model}
\par The membership inference attacks is to determine whether a data record is in a data set. Specifically, there is a target data set with sensitive information and a differential privacy mechanism is used to protect the target data set. Attackers has black-box access to the differential privacy mechanism. Given a data record, attackers guess: whether the given record is in the target data set. The formal definition of membership inference attacks is as follows:


\par \textbf{Definition 5} (membership inference attacks): Given black-box access to a differential privacy mechanism $M$ and a target record $x$, attackers guess whether the target record is in the the target data set $D$. 

$$ A(x,M)=\left\{
\begin{array}{rcl}
1      &      & {x\in D}\\
0     &      & {x\notin D}
\end{array} \right. $$

\par Here, the $A$ represents membership inference attacks method. If the attacks method asserts that the target record is in the target data set, it outputs 1. Otherwise, it outputs 0. 

\par There are three assumptions about the black-box access of the differential privacy mechanism. 
\par (1) For the same query, the black-box access returns the same answer no matter how many times the query is issued. If attackers can get a new answer for the same query from the black-box access when attackers issue the query every time, then attackers can get multiple i.i.d. samples of the target query's answer trivially. With multiple i.i.d. samples, it is trivial for attackers to infer sensitive information of records. Thus, this assumption is reasonable.

\par (2) For every different query, the black-box access returns an answer. The second assumption guarantees the utility of differential privacy mechanisms. If the  black-box access refuses to answer queries which are issued first time, the utility of differential privacy mechanisms is damaged, resulting in useless mechanisms.

\par (3) There is a threshold of the totally consumed privacy budget. When the totally consumed privacy budget is greater than the threshold, the black-box access aborts. The consumed privacy budget is accumulative and when totally consumed privacy budget is greater than a certain limitation, leaking information is trivial.


\par The ability of attackers is limited. The black-box access means that attackers issue a query to a differential privacy mechanism and gets a result as return.  Attackers cannot get any other information from the differential privacy mechanism.

\subsection {Security Analysis of The Laplace Mechanism}

\par The security of the Laplace mechanism is based on perturbation. The differential privacy requires that presence or absence of one record cannot be identified through mechanisms' outputs. The Laplace mechanism satisfies the requirement by perturbation. Specifically, queries' answer are perturbed by noise. The noise is drawn from a Laplace distribution whose variance is related to two factors. The first factor is strength of privacy guarantee. The second factor is maximum difference of query's answer, which is caused by presence or absence of any one record. The maximum difference is captured by the conception of sensitivity which is denoted by $\Delta D$. The strength of privacy guarantee is captured by the conception of privacy budget which is denoted by $\epsilon$. The variance of noise distribution is related to ratio of the sensitivity to the privacy budget. Specifically, the variance is $2(\Delta D/\epsilon)^2$.

\par The key idea behind the Laplace mechanism is: If the sensitivity is far less than fluctuation of added noise, it is hard to tell that variation of the Laplace mechanism's outputs happens due to noise distribution's fluctuation or due to presence or absence of a record. This idea can be described intuitively by formula $\Delta D \ll 2(\Delta D/\epsilon)^2$. 

\par The goal of covering the sensitivity can be achieved if the privacy budget is appropriately small. That is, for appropriately small privacy budget, the sensitivity could be much less than variance of noise distribution. 

\par In a word, for the Laplace mechanism, the sensitivity is covered by fluctuation of noise. So, the security foundation of the Laplace mechanism can be described by formula $\Delta D \ll 2(\Delta D/\epsilon)^2$ intuitively. However, the formula may not hold for some queries, which we will discuss in next subsection.

\par The hypothesis test is an appropriate tool to construct membership inference attacks method against the Laplace mechanism. In the Laplace mechanism, the difference caused by presence or absence of one record is covered by the fluctuation of noise. However, the hypothesis test method is designed to determine that variation  are just due to random noise or other reasons. In other words, the hypothesis test can be used to determine that variation of the Laplace mechanism's outputs is just due to randomness of noise or due to presence or absence of a record.    

\subsection { Counting Query }

\par The counting query is an appropriate query to perform membership inference attacks against the Laplace mechanism. In specific, it is acceptable for most cases that value of the privacy budget is from 0 to 1. Let's check the inequality $\Delta D \ll 2(\Delta D/\epsilon)^2$. For counting query, when the privacy budget is 1, the sensitivity is 1 and the variance of noise distribution is 2. The sensitivity is just a half of variance. When privacy budget is 0.5, the sensitivity is 1 and the variance is 8. So the sensitivity is one eighth of variance. In a word, the inequality $\Delta D \ll 2(\Delta D/\epsilon)^2$ may not hold for counting query when the privacy budget takes values which are generally considered to be an appropriate value for the  privacy budget.


\par There are other three reasons why the target linear query should be the counting query. Firstly, the counting query makes the attacks method technically right. The counting query satisfies linear property and the sensitivity of counting query is always 1. These two properties make it possible to obtain multiple i.i.d. samples for the counting query's answer. In addition, the hypothesis test method works on i.i.d. samples. Thus, the proposed membership inference attacks method is technically right due to unique properties of the counting query.

\par Secondly, the proposed attacks method is based on counting query so that it can work in a wide range. The counting query is very common. On the one hand, it is a common need to confirm the number of records which meet specific requirements, such as $sql$ statement " select count(*) from table student where student.age $\le$ 20 ". On the other hand, there are a lot of complex queries which are based on the counting query. 

\par Thirdly, the difference of query's answer, which is caused by presence or absence of one record, is most easily detected by hypothesis test method when the target query is the counting query. As mentioned before, the difference is smaller than or equal to the sensitivity and the sensitivity is far smaller than variance of noise distribution. Thus, it is hard to detect whether variation of the Laplace mechanism's outputs is due to presence or absence of one record. However, when the target  query is the counting query, the difference caused by presence or absence of any record is 1, reaching the value of sensitivity. So, the counting query makes it most easy to detect whether the variation of outputs is caused by presence or absence of one record.     


\subsection{Instance Attacks}

\par This subsection presents how to perform membership inference attacks against the Laplace mechanism. 

\par Attackers aim for determining whether a target record $x$ is in a target data set $D$. The background knowledge of attackers is denoted by $D_{know}$ which is a subset of target data set $D$. The target linear query is the counting query $q(\{x\}\cup D_{know})$. The $s$ represents the condition such that $D_s= \{x\}\cup D_{know}$. So, the target query can be expressed as $q_s(D)$ and its answer can be expressed as $M(q_s,D,\epsilon)$.	         
\par The membership inference attacks method consists of two subroutines. The first subroutine is to obtain multiple i.i.d. samples for the target query's answer. The first subroutine is feasible due to the linear property of the counting query, which is proved in previous section. By the first subroutine, attackers can obtain $m$ i.i.d. samples of the target query's answer and the samples are $\hat{a}_1,\hat{a}_2,\dots,\hat{a}_m$. When the target record $x$ is in the target data set $D$, 

\begin{eqnarray*}	
	q_s(D) &=& q(D_s\cap D) \\
	&=& q((D_{know}\cup\{x\})\cap D)\\
	&=& q(D_{know}\cup\{x\}) 
\end{eqnarray*}

\par Therefore, the mean of samples is $q(D_{know}\cup\{x\})$. When the target record $x$ is not in data set $D$
\begin{eqnarray*}	
	q_s(D) &=& q(D_s\cap D) \\
	&=& q((D_{know}\cup\{x\})\cap D)\\
	&=& q(D_{know}) 
\end{eqnarray*} 
\par Therefore, the mean of samples is $q(D_{know})$. 
\par Attackers can determine whether the target record is in the target data set by comparing $q(D_{know})$ with the mean of samples. The procedure of comparison is implemented by hypotheses test method in the second subroutine. 

\par In the second subroutine, the hypothesis test method is used to determine whether the samples which are obtained in the first subroutine are drawn from a distribution with mean $q(D_{know})$. If the mean of samples is determined to be $q(D_{know})$, the assertion is that the target record is not in the target data set. Otherwise, the assertion is that the target record is in the target data set.  

\par In the second subroutine, the key  is to find an appropriate hypothesis test method. That is, it is key to find an appropriate statistics. The one-sample t-test is chosen. The $T$ statistics follows $T$ distribution and the $T$ statistics is as follows 


\begin{eqnarray*}
	T &=& \frac{\bar{X} - \mu}{ S/\sqrt{m}}  
\end{eqnarray*}

\par Here, the $\mu$ is distribution mean and the $m$ is the number of samples. The $\bar{X}$ is the sample mean and the $S$ is sample standard variance. When the samples are denoted by $\hat{a}_1,\hat{a}_2,\dots,\hat{a}_m$ we have
\begin{eqnarray*}
	\bar{X} =  (\hat{a}_1+\hat{a}_2+\dots+\hat{a}_m)/m  
\end{eqnarray*}
\par And, 
\begin{eqnarray*}
	S=\sqrt{\frac{(\hat{a}_1-\bar{X})^2+(\hat{a}_2-\bar{X})^2+\dots+(\hat{a}_m-\bar{X})^2}{m}}   \\
\end{eqnarray*}

\par The one-sample t-test is an appropriate hypothesis test method because it is designed to determine whether some samples are drawn from a distribution with a certain mean.  

\par The significance level is 0.05. There are two widely accepted choices for significance level, namely $0.01$ and $0.05$. We desire that as long as difference is statistically significant, the alternative hypothesis could be accepted. So we choose $0.05$ as value of significance level.

\par There is a table where $p$ value can be searched by the freedom and the value of statistic $T$. The freedom is $m-1$ and the value of statistic $T$ can be calculated as above.

\begin{algorithm}
	\caption*{Membership Inference Attacks Method A}
	\label{alg:Framwork} 
	\begin{algorithmic}[1]
		\REQUIRE ~~\\ 
		the background knowledge $D_{know}$ of attackers;\\		 
		the target record $x$;\\
		the Laplace mechanism $M$;\\
		the counting  query $q$;\\
		total privacy budget $\epsilon_t$;\\		 
		the number of samples $m$;\\	 
		the significance level $\alpha=0.05$;\\
		
		\ENSURE ~~\\
		the $x$ is in data set $D$ or not;\\
		
		\STATE construct conditions $s$ such that $D_s = D_{know} \cup \{x\}$; 
		\STATE let $\epsilon = \frac{\epsilon_t}{m}$;		
		\FOR{$i=1$ to $m$}
		\STATE $\hat{a}_i = M(q_s,D,\epsilon)$ by the first subroutine;
		\ENDFOR
		
		\STATE make\ hypothesis: \\
		the null hypothesis is $H_0$: $\mu_0 = q(D_{know})$; \\
		the alternative hypothesis is $H_1$: $\mu_0 \ne q(D_{know})$; \\
		\STATE calculate $\bar{X} = \frac{\hat{a}_1+\hat{a}_2+\dots+\hat{a}_m}{m}$;\\
		\STATE calculate $S=\sqrt{\frac{(\hat{a}_1-\bar{X})^2+(\hat{a}_2-\bar{X})^2+\dots+(\hat{a}_m-\bar{X})^2}{m}}$;\\
		\STATE calculate $T = \frac{\bar{X} - \mu_0}{\frac{S}{\sqrt{m}}}$;\\
		\STATE find $p$ according to $T$ and freedom $m-1$;\\
		if $p<\alpha$, the assertion is $x\in$ D; \\
		if $p>\alpha$, the assertion is $x\notin D$ \\ 
		\RETURN assertion

	\end{algorithmic}
\end{algorithm}

\subsection{Success Rate Analysis}

\par The success rate of proposed attacks method is tightly related to two factors including the amount of background knowledge of attackers and the threshold of privacy budget in black-box access. According to the two factors, there are four cases faced by attackers,  as shown in Fig 2.

\begin{figure}[htbp]
	\centering
	\includegraphics[width=7.0cm]{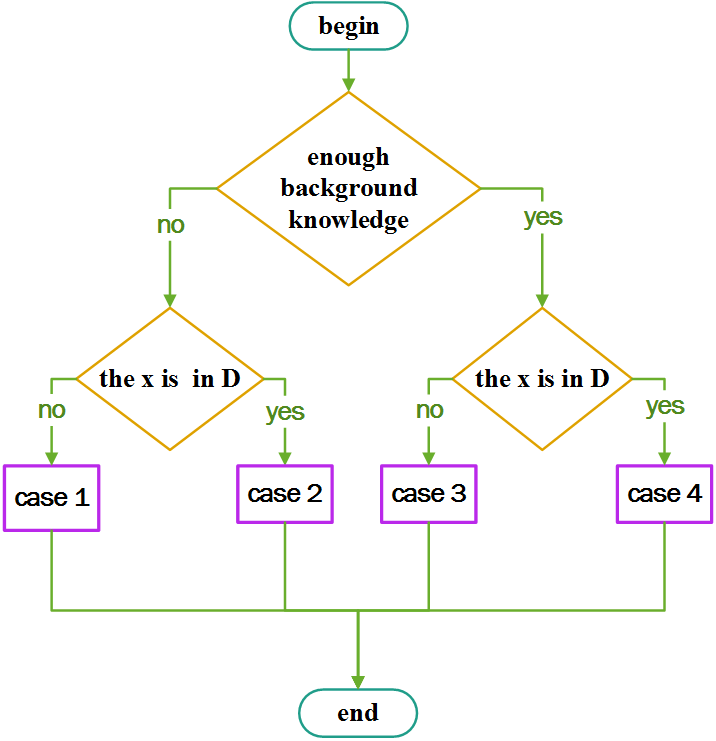}
	\caption*{Fig 2: cases faced by attackers} 
\end{figure}

\par Suppose that the threshold of the privacy budget is $\epsilon_l$. When the totally consumed privacy budget is greater than the threshold from the differential privacy mechanism's perspective, the black-box access of differential privacy mechanisms aborts. The "enough background knowledge" means that attackers can construct as many as needed number of queries such that the totally consumed privacy budget is greater than the threshold, resulting in abortion of black-box access. 


\par For the case 4, there is enough background knowledge available to attackers. That is, there is enough number of records in $D_{know}$. In addition, the target record $x$ is in the target data set. By the theorem 4, from the differential privacy mechanism's perspective, the totally consumed privacy budget is $m\epsilon$. Because the background knowledge is enough, attackers can make the $m$ big enough such that $m\epsilon>\epsilon_l$, result in abortion of the black-box. 

\par For the case 3, there is enough number of records in $D_{know}$ and the target record $x$ is in not the target data set. By the theorem 4, from the differential privacy mechanism's perspective, the totally consumed privacy budget is $\epsilon$ no matter how many queries attackers issue. So the black-box access never aborts. By the theorem 3, from attackers' perspective, the totally consumed privacy budget is $m\epsilon$. Thus, attackers can issue as many as needed queries such that the $m\epsilon$ is unreasonably large, resulting in that attackers can find the absence of the target record trivially by the hypothesis test method.   

\par For the case 2 and the case 1, the background knowledge is not enough. Thus, attackers cannot consume unreasonably large privacy budget  or force the black-box access to abort. In these cases, the success rate is tightly related to the exact amount of background knowledge. For these two cases, we will give a delicate analysis about the success rate in theorem 5.

\par In brief, from attackers' perspective, they issue as many as possible queries to consume as much as possible privacy budget. In specific, there are three possible results. Firstly, the black-box access aborts, which means that the totally consumed privacy budget is greater than the threshold and the target record is in the target data set. Secondly, the black-box access does not aborts and the totally consumed privacy budget is so big that attackers can find the absence of the target record. Thirdly, all records of attackers' background knowledge are used and attackers obtain multiple answers of the target query. Attackers determine whether the target record is in the target data set by hypothesis test method. The success rate of using hypothesis test method is analyzed in theorem 5 at next. 

\begin{table*}[!htbp]
	\centering
	\begin{tabular}{|c|c|c|c|}
		\hline
		\diagbox{assertion}{hypothesis}& $H_0$& $H_1$\\ 
		\hline
		$H_0$& $P\{accetp\ H_0|H_0\ is\ ture\} = 1-\alpha = 0.95 $& $P\{accetp\ H_0|H_1\ is\ ture\} = \delta $\\
		\hline
		$H_1$& $P\{accetp\ H_1|H_0\ is\ ture\} = \alpha = 0.05 $ & $P\{accetp\ H_1|H_1\ is\ ture\} = 1-\delta $\\
		\hline		 
	\end{tabular}
	\caption*{Table 1: success rate of hypothesis test}
\end{table*}

\par The hypothesis test method has two types error shown in Table 1. Based on the Table 1, the success rate of proposed attacks method can be calculated. The proof of theorem 5 is deferred to the appendix.
\par \textbf{Theorem 5} The success rate $R$ of the proposed membership inference attacks method is 

\begin{eqnarray*}
	R = \frac{1}{2}(1.95-\int_{-T_{(m-1,0.975)}+\frac{(\mu_0-\mu_1)*\sqrt{m}}{S}}^{T_{(m-1,0.975)}+\frac{(\mu_0-\mu_1)*\sqrt{m}}{S}}f(t)dt)\\
\end{eqnarray*}
$\hfill\Box$

\par Here, the $T_{(m-1,0.975)}$ is the value of $T$ statistic when the freedom of $T$ distribution is $m-1$ and the cumulative probability is 0.975. The $f(t)$ is the probability density function of $T$ distribution with freedom $m-1$. The $\mu_0$ is the mean assumed in null hypothesis. The $\mu_1$ is the true mean of distribution when the null hypothesis is wrong. We have 


\begin{eqnarray*}
	\mu_0= q(D_{know}) \ \ and \ \ \mu_1 = q(D_{know}\cup\{x\})
\end{eqnarray*}

\par  In above formula, $\mu_0-\mu_1$ is equal to $-1$ because the query $q$ is the counting query. The $m$ is the number of obtained samples from the Laplace mechanism by black-box access. The $S$ is the standard deviation of obtained samples. 

\par As mentioned before, the standard deviation of noise distribution is $\sqrt{2}\Delta D/\epsilon$. In addition, for the counting query the sensitivity $\Delta D$ is 1. The standard deviation of sample could be substituted with standard deviation of noise distribution. That is, the $S$ could be substituted with $\sqrt{2}\Delta D/\epsilon$. Then, the $R$ is
\begin{eqnarray*}
	R \approx \frac{1}{2}(1.95-\int_{-T_{(m,0.975)} -\frac{\epsilon\sqrt{m}}{\sqrt{2}}}^{T_{(m,0.975)}-\frac{\epsilon\sqrt{m}}{\sqrt{2}}}f(t)dt)\\
\end{eqnarray*}

\par The $\epsilon$ is the privacy budget consumed by each query. By the above formula, the $R$ increases when the number of obtained samples increases. 
\par In terms of totally consumed privacy budget, the pattern is different. In specific, the totally consumed privacy budget $\epsilon_t$ is equal to $m\epsilon$. So, the $R$ is

\begin{eqnarray*}
	R \approx \frac{1}{2}(1.95-\int_{-T_{(m,0.975)} -\epsilon_t/\sqrt{2m}}^{T_{(m,0.975)}-\epsilon_t/\sqrt{2m}}f(t)dt)\\
\end{eqnarray*}
\par According to the above formula, the $R$ decreases when the number of obtained samples increases under the condition that the totally consumed privacy budget is predetermined.

\par By the theorem 2, the maximum number of samples is equal to the number of records in $D_{know}$ at most. Let $r$ be the number of records in $D_{know}$. The above two formulas become  
\begin{eqnarray*}
	R \approx \frac{1}{2}(1.95-\int_{-T_{(r,0.975)} -\frac{\epsilon\sqrt{r}}{\sqrt{2}}}^{T_{(r,0.975)}-\frac{\epsilon\sqrt{r}}{\sqrt{2}}}f(t)dt)\\
\end{eqnarray*}
\par And
\begin{eqnarray*}
	R \approx \frac{1}{2}(1.95-\int_{-T_{(r,0.975)} -\epsilon_t/\sqrt{2r}}^{T_{(r,0.975)}-\epsilon_t/\sqrt{2r}}f(t)dt)\\
\end{eqnarray*}

\par In above analyses, the consumed privacy budget is calculated from attackers' perspective. However, there are two cases if the consumed privacy budget  is calculated from differential privacy mechanisms' perspective. Firstly, if the target record is in the target data set, the consumed privacy budget is the same with that of attackers' perspective. Secondly, if the target record is not in the target data set, the consumed privacy budget is $\epsilon$ which is the amount of privacy budget consumed by one query.

\section{experiment}
\par In this section, the detailed information of extensive experiments is given. The goal is to evaluate the success rate of proposed membership inference attacks method. As analyzed before, for the case 3 and 4, attackers can determine that the target record is in the target data set or not with probability 1. In this section, we evaluate the success rate for the case 1 and 2 where the background knowledge of attackers are not enough.  

\par For the case 1 and 2, the success rate is related to two factors, namely consumed privacy budget and the number of obtained samples. In specific, the privacy budget is the key parameter to control the amount of information released by differential privacy mechanisms. The number of samples has important influence on performance of hypothesis test method. The intuition is consistent with the theoretical analysis in theorem 5. So we pay our attention to the two factors and investigate their influence on the success rate. All experiments are based on the counting query. 


\par The privacy budget is unreasonably large if the value of privacy budget is 10 in real application scenarios. However, the key contribution of this paper is that the totally consumed privacy budget is different for delicately designed queries from attackers' perspective and from differential privacy mechanisms' perspective. As analyzed before, attackers could consume unreasonably large privacy budget while the differential privacy mechanism doesn't beware it. This is the reason why we will choose a large value for totally consumed privacy budget to verify the success rate of proposed attacks method.

\par The number of samples is from 4 to 29. When the number of samples is less than 4, the success rate is not good enough. The reason is that the number of samples are so small that the standard deviation of samples has big error. For example, if there is only one sample, then the sample standard deviation is 0, resulting in huge error. According to experimental results, when the number of samples is less than 4, there are big fluctuation in experimental results. In order to show patterns of experimental results, experiments need to be repeated so many times if the number of samples is less than 4.

\begin{figure*}[htbp]
	\centering 
	\begin{minipage}[t]{0.45\linewidth}
		\centering
		\includegraphics[width=5.7cm]{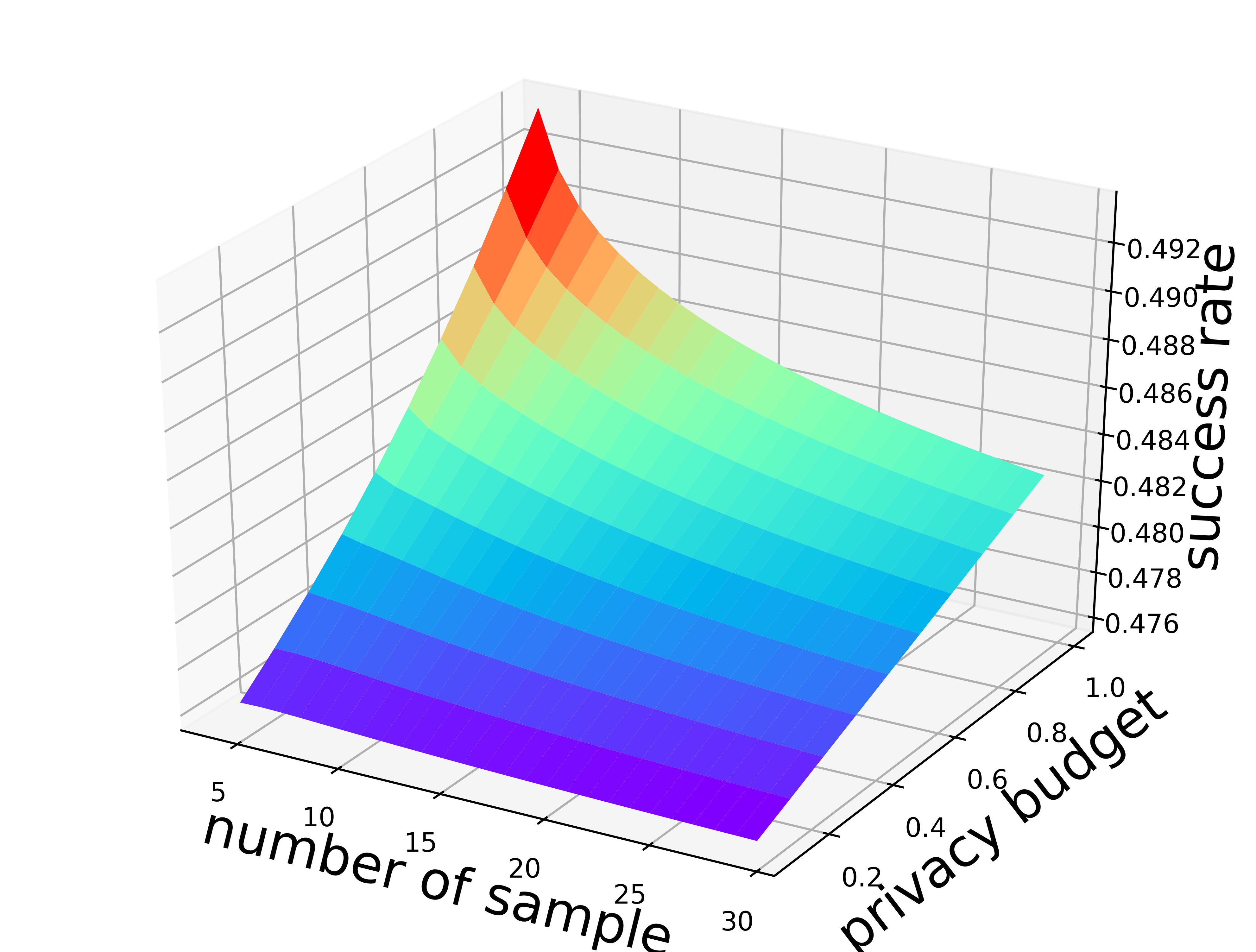}
		\caption*{(a)theoretical results\\  privacy budget is from 0.1 to 1  }
	\end{minipage}
	\begin{minipage}[t]{0.45\linewidth}
		\centering
		\includegraphics[width=5.7cm]{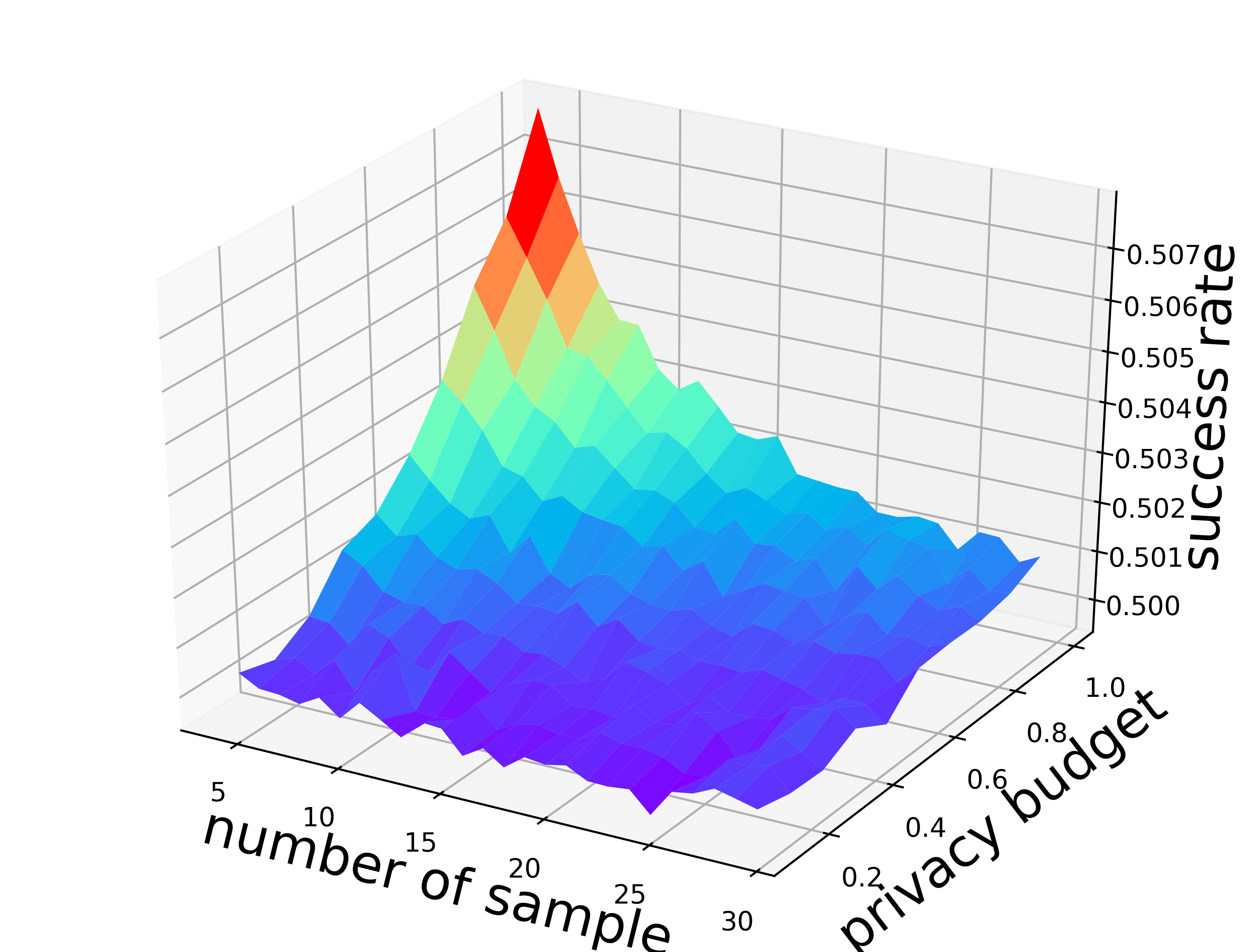}
		\caption*{(b)experiment results\\ privacy budget is from 0.1 to 1}
	\end{minipage}
	\begin{minipage}[t]{0.45\linewidth}
		\centering
		\includegraphics[width=5.7cm]{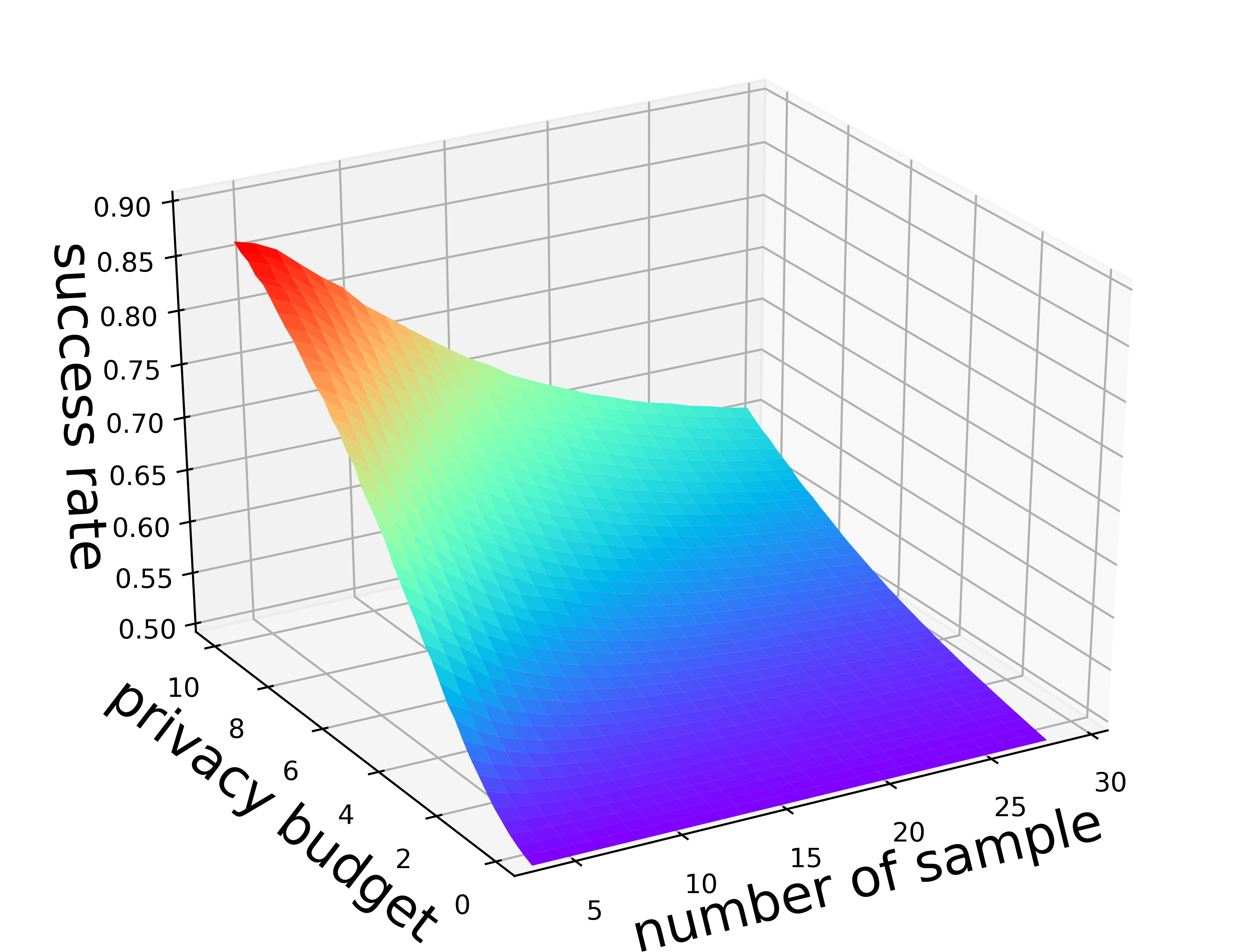}
		\caption*{(c)theoretical results\\  privacy budget is from 0.1 to 10}
	\end{minipage}
	\begin{minipage}[t]{0.45\linewidth}
		\centering
		\includegraphics[width=5.7cm]{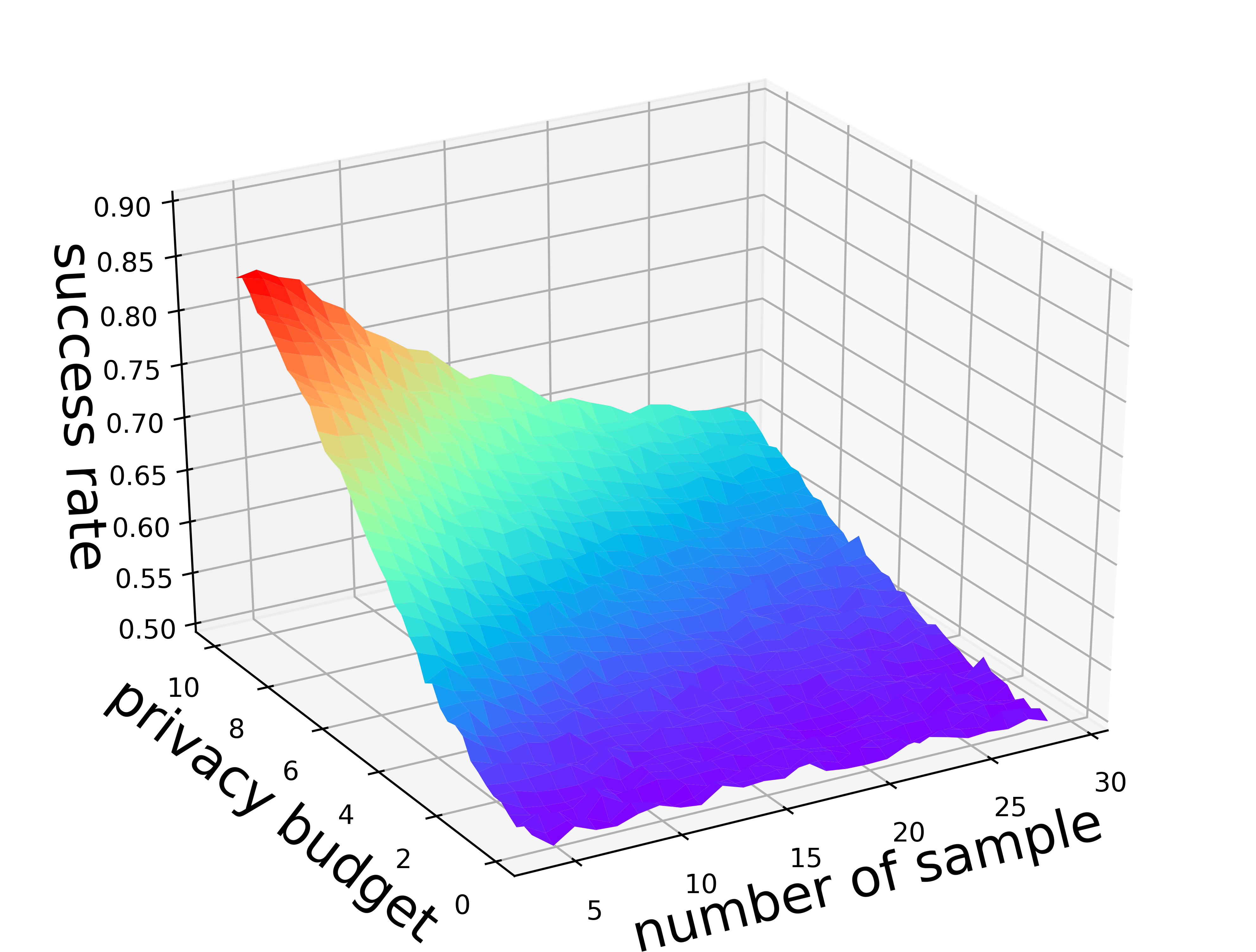}
		\caption*{(d)experiment results\\ privacy budget is from 0.1 to 10}
	\end{minipage}	 	 
	\caption*{Fig 3: totally consumed privacy budget is a constant }
\end{figure*}

\begin{figure*}[htbp]
	\centering 
	\begin{minipage}[t]{0.45\linewidth}
		\centering
		\includegraphics[width=5.7cm]{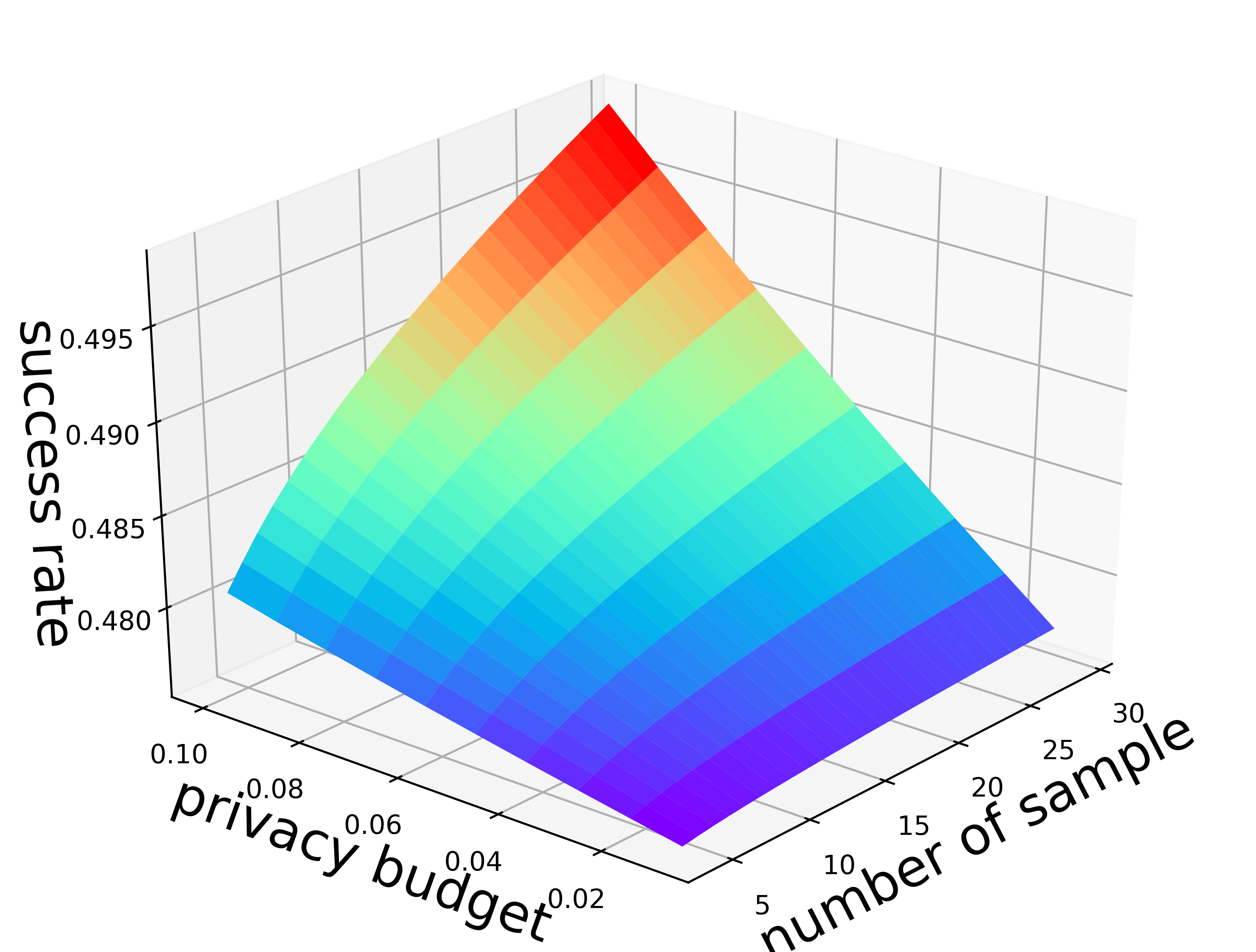}
		\caption*{(a)theoretical results\\ privacy budget is from 0.01 to 0.1}
	\end{minipage}
	\begin{minipage}[t]{0.45\linewidth}
		\centering
		\includegraphics[width=5.7cm]{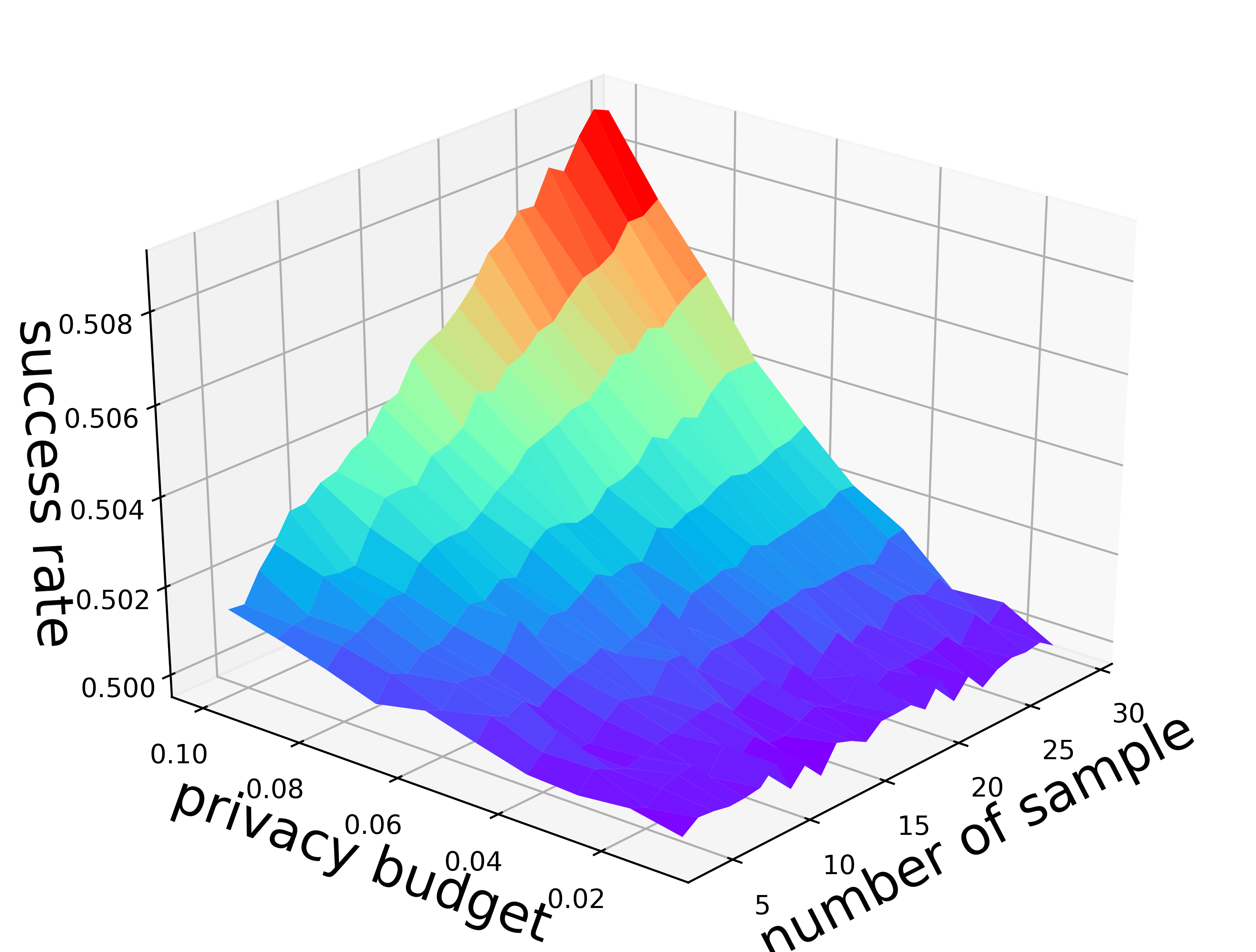}
		\caption*{(b)experiment results\\ privacy budget is from 0.01 to 0.1}
	\end{minipage}
	\begin{minipage}[t]{0.45\linewidth}
		\centering
		\includegraphics[width=5.7cm]{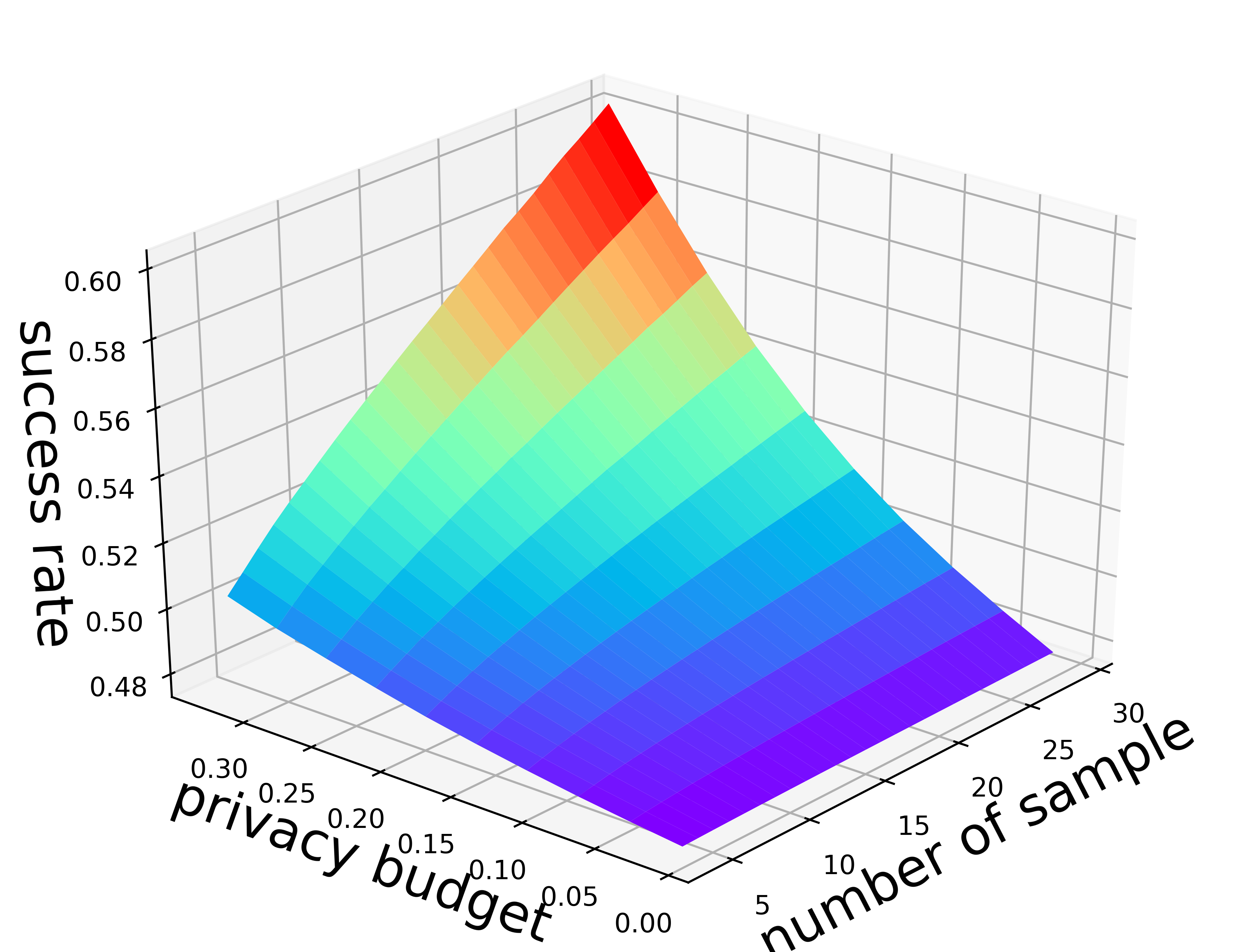}
		\caption*{(c)theoretical results\\ privacy budget is from 0.01 to 0.33}
	\end{minipage}
	\begin{minipage}[t]{0.45\linewidth}
		\centering
		\includegraphics[width=5.7cm]{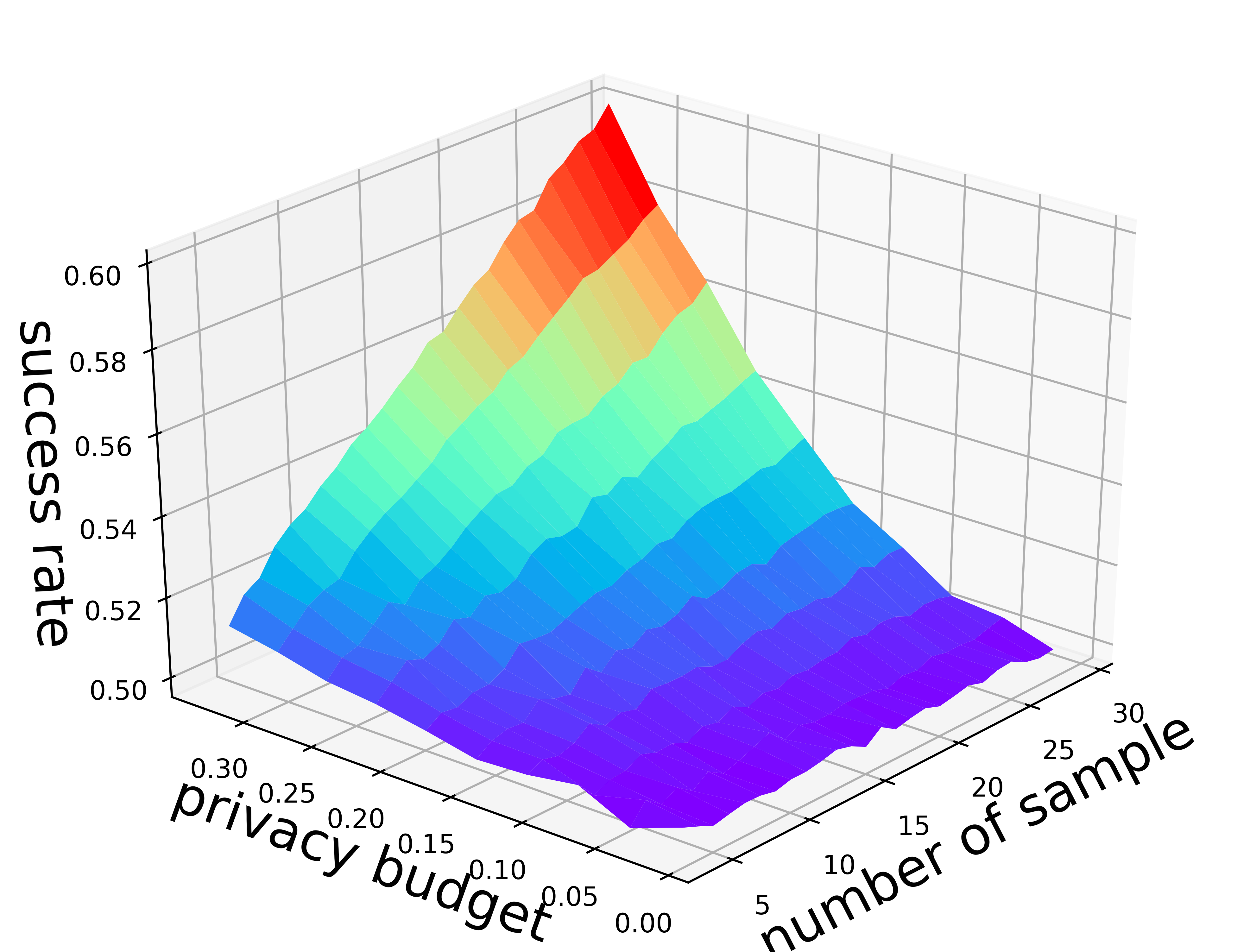}
		\caption*{(d)experiment results\\ privacy budget is from 0.01 to 0.33}
	\end{minipage}	 	 
	\caption*{Fig 4: consumed privacy budget of each query is a constant }
\end{figure*}

\par  In the first experiment, suppose that the totally consumed privacy budget is a constant to verify the theoretical analysis of the success rate. The privacy budget is from 0.1 to 1 and from 0.1 to 10 respectively. The number of samples is from 4 to 29.  We repeat experiments and show average of these experimental results in Fig 3.           

\par In Fig 3, the privacy budget is the totally consumed privacy budget by all samples. In specific, if the privacy budget is 1 and the number of samples is 10, consumed privacy budget of each sample is 0.1.


\par According to results in Fig 3, the experimental results are consistent with theoretical analysis. The success rate increases when the totally consumed privacy budget increases. If the totally consumed privacy becomes bigger, the privacy budget allocated to each sample becomes bigger. Thus, each sample is perturbed with smaller noise. So, the hypothesis test method has better performance. When the totally consumed privacy budget is constant, the success rate decreases as the number of samples increases.  The reason is that when the number of samples is bigger the privacy budget of every sample is smaller, resulting in bigger noise of each sample. For example, totally consumed privacy budget is 1. When the number of samples is 5, the privacy budget for every sample is 0.2. While when the number of samples is 10, the privacy budget for every sample is 0.1.

\par In the second experiment, suppose that privacy budget of each sample is constant to verify theoretical analysis of the success rate. As in the first experiment, the range of the number of samples is from 4 to 29. The privacy budget is from 0.01 to 0.1 and from 0.01 to 0.33 respectively. We repeat experiments and show average of these experimental results in Fig 4.

\par In Fig 4, the privacy budget is the consumed privacy budget of each sample. In specific, if the privacy budget is 0.2 and the number of samples is 10, consumed privacy budget  is 2 in total.  

\par According to results in Fig 4, the experimental results are consistent with theoretical analysis. The success rate increases when the privacy budget of each sample increases. When the privacy budget is smaller, the added noise is bigger. It is hard to tell that variation of Laplace mechanism's outputs is due to presence or absence of one record or due to fluctuation of added noise. So the success rate is low. When the privacy budget is bigger, the added noise is smaller so it is easier to tell the variation caused by presence or absence of one record from fluctuation of added noise.

\par The success rate increases when the number of samples increases. Intuitively speaking, when the privacy budget of each sample is constant, the amount of information of each sample is almost the same. So, when the number of samples increases, the amount of information released by differential privacy mechanisms increases, resulting in increase of the success rate.        

\par In all above experiments, the totally consumed privacy budget is calculated from attackers' perspective. Because the proposed method is a method to perform attacks, it is meaningful to calculate totally consumed privacy budget from attackers' perspective.  

\par From differential privacy mechanisms' perspective, if the totally consumed privacy budget exceeds the threshold of the privacy budget, the black-box access aborts, resulting in that attackers know presence of the target record in the target data set. If the totally consumed privacy budget does not exceed the threshold , attackers can continually consume more privacy budget until the success rate is acceptable for attackers or all records in $D_{know}$ are used.

\section{conclusion}

\par In this paper, we find that the differential privacy does not take liner property of queries into account, resulting in unexpected information leakage. In specific, the totally consumed privacy budget of delicately designed queries may be different from attackers' perspective and from differential privacy mechanisms' perspective due to linear property. The difference leads to unexpected information leakage. In addition, we show how attackers can consume extra privacy budget by its background knowledge, which is against the old opinion that the amount of background knowledge almost has no influence on privacy guarantee of differential privacy.  

\par In order to demonstrate the unexpected information leakage, we show a membership inference attacks against the Laplace mechanism. Specifically, based on linear property of queries, a method is proposed to obtain multiple i.i.d. samples of the linear query's answer. Based on these samples, the hypothesis test method is used to determine whether a target record is in a target data set. Based on counting query, we perform extensive experiments to verify proposed membership inference attacks method. According to experimental results, the proposed membership inference attacks method works well.

\bibliographystyle{plain}
\bibliography{acmart}

\begin{thebibliography}{10}

\bibitem{37}
Nour Almadhoun, Erman Ayday, and {\"O}zg{\"u}r Ulusoy.
\newblock Differential privacy under dependent tuples?the case of genomic
  privacy.
\newblock {\em Bioinformatics}, 36(6):1696--1703, 2020.

\bibitem{38}
Nour Almadhoun, Erman Ayday, and {\"O}zg{\"u}r Ulusoy.
\newblock Inference attacks against differentially private query results from
  genomic datasets including dependent tuples.
\newblock {\em Bioinformatics}, 36(Supplement\_1):i136--i145, 2020.

\bibitem{44}
Michael Backes, Pascal Berrang, Mathias Humbert, and Praveen Manoharan.
\newblock Membership privacy in microrna-based studies.
\newblock In {\em Proceedings of the 2016 ACM SIGSAC Conference on Computer and
  Communications Security}, CCS 2016, pages 319--330, New York, NY, USA, 2016.
  Association for Computing Machinery.

\bibitem{35}
J.~{Chen}, H.~{Ma}, D.~{Zhao}, and L.~{Liu}.
\newblock Correlated differential privacy protection for mobile crowdsensing.
\newblock {\em IEEE Transactions on Big Data}, pages 1--1, 2017.

\bibitem{29}
Damien Desfontaines and Bal{\'a}zs Pej{\'o}.
\newblock Sok: Differential privacies.
\newblock {\em Proceedings on Privacy Enhancing Technologies},
  2020(2):288--313, 2020.

\bibitem{30}
Weibei Fan, Jing He, Mengjiao Guo, Peng Li, Zhijie Han, and Ruchuan Wang.
\newblock Privacy preserving classification on local differential privacy in
  data centers.
\newblock {\em Journal of Parallel and Distributed Computing}, 135:70--82,
  2020.

\bibitem{27}
Quan Geng, Wei Ding, Ruiqi Guo, and Sanjiv Kumar.
\newblock Tight analysis of privacy and utility tradeoff in approximate
  differential privacy.
\newblock In {\em International Conference on Artificial Intelligence and
  Statistics}, pages 89--99, 2020.

\bibitem{36}
Nils Homer, Szabolcs Szelinger, Margot Redman, David Duggan, Waibhav Tembe,
  Jill Muehling, John~V Pearson, Dietrich~A Stephan, Stanley~F Nelson, and
  David~W Craig.
\newblock Resolving individuals contributing trace amounts of dna to highly
  complex mixtures using high-density snp genotyping microarrays.
\newblock {\em PLoS genetics}, 4(8), 2008.

\bibitem{26}
Wen Huang, Shijie Zhou, Yongjian Liao, and Hongjie Chen.
\newblock An efficient differential privacy logistic classification mechanism.
\newblock {\em IEEE Internet of Things Journal}, 6(6):10620--10626, 2019.

\bibitem{28}
Wen Huang, Shijie Zhou, Yongjian Liao, and Ming Zhuo.
\newblock Optimizing query times for multiple users scenario of differential
  privacy.
\newblock {\em IEEE Access}, 7:183292--183299, 2019.

\bibitem{22}
Noah Johnson, Joseph~P Near, and Dawn Song.
\newblock Towards practical differential privacy for sql queries.
\newblock {\em Proceedings of the VLDB Endowment}, 11(5):526--539, 2018.

\bibitem{39}
Yang Li, Dasen Yang, and Xianbiao Hu.
\newblock A differential privacy-based privacy-preserving data publishing
  algorithm for transit smart card data.
\newblock {\em Transportation Research Part C: Emerging Technologies},
  115:102634, 2020.

\bibitem{33}
Denglong Lv and Shibing Zhu.
\newblock Correlated differential privacy protection for big data.
\newblock In {\em 2018 IEEE 32nd International Conference on Advanced
  Information Networking and Applications (AINA)}, pages 1011--1018. IEEE,
  2018.

\bibitem{23}
Kobbi Nissim, Sofya Raskhodnikova, and Adam Smith.
\newblock Smooth sensitivity and sampling in private data analysis.
\newblock In {\em Proceedings of the thirty-ninth annual ACM symposium on
  Theory of computing}, pages 75--84, 2007.

\bibitem{41}
Md~Atiqur Rahman, Tanzila Rahman, Robert Lagani{\`e}re, Noman Mohammed, and
  Yang Wang.
\newblock Membership inference attack against differentially private deep
  learning model.
\newblock {\em Transactions on Data Privacy}, 11(1):61--79, 2018.

\bibitem{43}
Reza Shokri, Marco Stronati, Congzheng Song, and Vitaly Shmatikov.
\newblock Membership inference attacks against machine learning models.
\newblock In {\em 2017 IEEE Symposium on Security and Privacy (SP)}, pages
  3--18. IEEE, 2017.

\bibitem{31}
Hao Wang, Zhengquan Xu, Shan Jia, Ying Xia, and Xu~Zhang.
\newblock Why current differential privacy schemes are inapplicable for
  correlated data publishing.
\newblock {\em World Wide Web}, pages 1--23.

\bibitem{32}
Genqiang Wu, Xianyao Xia, and Yeping He.
\newblock Extending differential privacy for treating dependent records via
  information theory.
\newblock {\em arXiv preprint arXiv}, 1703, 2017.

\bibitem{19}
Xiaotong Wu, Taotao Wu, Maqbool Khan, Qiang Ni, and Wanchun Dou.
\newblock Game theory based correlated privacy preserving analysis in big data.
\newblock {\em IEEE Transactions on Big Data}, 2017.

\bibitem{25}
Bin Yang, Issei Sato, and Hiroshi Nakagawa.
\newblock Bayesian differential privacy on correlated data.
\newblock In {\em Proceedings of the 2015 ACM SIGMOD international conference
  on Management of Data}, pages 747--762. ACM, 2015.

\bibitem{42}
Samuel Yeom, Irene Giacomelli, Matt Fredrikson, and Somesh Jha.
\newblock Privacy risk in machine learning: Analyzing the connection to
  overfitting.
\newblock In {\em 2018 IEEE 31st Computer Security Foundations Symposium
  (CSF)}, pages 268--282. IEEE, 2018.

\bibitem{34}
Tao Zhang, Tianqing Zhu, Ping Xiong, Huan Huo, Zahir Tari, and Wanlei Zhou.
\newblock Correlated differential privacy: feature selection in machine
  learning.
\newblock {\em IEEE Transactions on Industrial Informatics}, 16(3):2115--2124,
  2019.

\bibitem{24}
Tianqing Zhu, Ping Xiong, Gang Li, and Wanlei Zhou.
\newblock Correlated differential privacy: Hiding information in non-iid data
  set.
\newblock {\em IEEE Transactions on Information Forensics and Security},
  10(2):229--242, 2014.

\end{thebibliography}
\section{appendix}
\par \textbf{Theorem 5} The success rate $R$ of the proposed membership inference attacks is equal to 

\begin{eqnarray*}	 
	R = \frac{1}{2}(1.95-\int_{-T_{(m,0.975)}+\frac{(\mu_0-\mu_1)*\sqrt{m}}{S}}^{T_{(m,0.975)}+\frac{(\mu_0-\mu_1)*\sqrt{m}}{S}}f(t)dt)\\
\end{eqnarray*}

\par \textbf{Proof}
\par As shown in table 1, there are two types of error of hypothesis test. According the Table 1, we have
\begin{eqnarray*}
	R&=&\frac{(1-\alpha)+(1-\delta)}{(1-\alpha) + \delta + \alpha + (1-\delta)} \\
	&=& \frac{1}{2}(2-\alpha-\delta)\\
	&=& \frac{1}{2}(1.95-\delta)\\
\end{eqnarray*}
\par For the $\alpha$, we have 

\begin{eqnarray*}
	& & \alpha \\
	&=& P\{accetp\ H_1|H_0\ is\ ture\}  \\
	&=& P\{\frac{\bar{X} - \mu_0}{\frac{S}{\sqrt{m}}}<-T^*|H_0\ is\ ture \} \\
	&+& P\{T^*<\frac{\bar{X} - \mu_0}{\frac{S}{\sqrt{m}}}|H_0\ is\ ture \} \\
	&=& 2*P\{T^*<\frac{\bar{X} - \mu_0}{\frac{S}{\sqrt{m}}}|H_0\ is\ ture \} \\
	&=& 2*(1-P\{\frac{\bar{X} - \mu_0}{\frac{S}{\sqrt{m}}}<T^*|H_0\ is\ ture \}) \\
	&=& 0.05	
\end{eqnarray*}
\par We have, 
\begin{eqnarray*}
	P\{ \frac{\bar{X} - \mu_0}{\frac{S}{\sqrt{m}}}<T^*|H_0\ is\ ture \} = 0.975
\end{eqnarray*}

\par Namely,
\begin{eqnarray*}
	P\{ \frac{\bar{X} - \mu_0}{\frac{S}{\sqrt{m}}}<T^*\} = 0.975
\end{eqnarray*}

\par The $\frac{\bar{X} - \mu_0}{\frac{S}{\sqrt{m}}}$ obeys $T$ distribution with freedom $m-1$. Let the $T^*$ be value of $T$ statistic with freedom $m-1$ and probability 0.975, namely $T^* = T_{(m,0.975)}$. In addition, let $f(t)$ be the probability density function of $T$ statistic with freedom $m-1$ and probability 0.975. We have

\begin{eqnarray*}
	\int_{-\infty}^{0}f(t)dt+ \int_{0}^{T^*}f(t)dt = 0.975 \\
\end{eqnarray*}
\par So, we have
\begin{eqnarray*}
	\int_{0}^{T^*}f(t)dt = 0.475
\end{eqnarray*}

\par For the $\delta$, we have
\begin{eqnarray*}
	& & \delta \\
	&=& P\{accetp\ H_0|H_1\ is\ ture\} \\
	&=& P\{-T^*< \frac{\bar{X} - \mu_0}{\frac{S}{\sqrt{m}}}<T^*|H_1\ is\ ture \} \\
\end{eqnarray*} 

\par Suppose that the true mean is $\mu_1$, we have
\begin{eqnarray*}
	& & \delta \\
	&=& P\{-T^*+\frac{\mu_0 - \mu_1}{\frac{S}{\sqrt{m}}} < \frac{\bar{X} - \mu_1}{\frac{S}{\sqrt{m}}}<T^*+\frac{\mu_0 - \mu_1}{\frac{S}{\sqrt{m}}}|H_1\ is\ ture \} \\
	&=& P\{-T^*+\frac{\mu_0 - \mu_1}{\frac{S}{\sqrt{m}}} < \frac{\bar{X} - \mu_1}{\frac{S}{\sqrt{m}}}<T^*+\frac{\mu_0 - \mu_1}{\frac{S}{\sqrt{m}}} \} \\
\end{eqnarray*}
\par Denote $\frac{\mu_0 - \mu_1}{\frac{S}{\sqrt{m}}}$ by $T_1$, we have
\begin{eqnarray*}
	\delta = \int_{-T^*+T_1}^{T^*+T_1}f(t)dt \\
\end{eqnarray*}
\par So, we have
\begin{eqnarray*}
	R&=& \frac{1}{2}(1.95-\delta)\\
	&=& \frac{1}{2}(1.95-\int_{-T^*+T_1}^{T^*+T_1}f(t)dt)\\
	&=& \frac{1}{2}(1.95-\int_{-T_{(m,0.975)}+\frac{(\mu_0-\mu_1)*\sqrt{m}}{S}}^{T_{(m,0.975)}+\frac{(\mu_0-\mu_1)*\sqrt{m}}{S}}f(t)dt)\\
\end{eqnarray*}

$\hfill\Box$

\end{document}